
\documentclass[12pt]{article}
\usepackage[latin9]{inputenc}
\usepackage{amsmath}
\usepackage{amssymb}
\usepackage[numbers,sort&compress]{natbib}
\usepackage{amsfonts}
\usepackage{indentfirst}
\usepackage{graphicx}
\usepackage{color}

\setcounter{MaxMatrixCols}{10}

\numberwithin{equation}{section}

\definecolor{darkgreen}{rgb}{0,0.35,0}

\providecommand{\U}[1]{\protect\rule{.1in}{.1in}}
\textheight 23cm\textwidth 17cm \oddsidemargin -5pt \evensidemargin
0pt \topmargin -50pt \hyphenation{sym-me-tri-za-tion}
{document}
\font\myfont=cmr12 at 22pt

\date{}
\title{{\myfont Generalizing the $\mathfrak{bms}_{3}$ and 2D-conformal
algebras by expanding the Virasoro algebra}}
\author{Ricardo Caroca$^{1}$\thanks{%
rcaroca@ucsc.cl}, \,\,Patrick Concha$^{2}$\thanks{%
patrick.concha@pucv.cl}, \\
Evelyn Rodr\'iguez$^{3}$\thanks{%
evelyn.rodriguez@edu.uai.cl} \,\,and\,\,Patricio Salgado-Rebolledo$^{4}$%
\thanks{%
patricio.salgado@uai.cl} \\
\\
[3pt] {\small {}{}{}$^{1}$}\textit{{\small {}{}{}Departamento de Matemática
y F\'isica Aplicadas }}{\small {}{}}\\
{\small {}{} \ }\textit{{\small {}{}{}Universidad Cat\'olica de la
Sant\'isima Concepci\'on}}{\small {}{}}\\
{\small {}{} \ {}Alonso de Rivera 2850, Concepción, Chile}\\
[15pt] {\small {}{}{}$^{2}$}\textit{{\small {}{}{}Instituto de F\'isica,
Pontificia Universidad Cat\'olica de Valpara\'iso}}{\small {}{}}\\
{\small {}{} \ {}Casilla 4059, Valpara\'iso, Chile}\\
[15pt] {\small {}{}{}$^{3}$}\textit{{\small {}{}{}Departamento de Ciencias,
Facultad de Artes Liberales}}{\small {}{}}\\
{\small {}{} \ }\textit{{\small {}{}{}Universidad Adolfo Ib\'añez}}{\small %
{}{}}\\
{\small {}{} {}Av. Padre Hurtado 750, Viña del Mar, Chile}\\
[15pt] $^{{\small 4}}$\textit{{\small {}{}{}Facultad de Ingenier\'ia y
Ciencias and UAI Physics Center}}{\small {}{}}\\
{\small {}{} }\textit{{\small {}{}{}Universidad Adolfo Ib\'añez}}{\small {}{}%
}\\
{\small {}{} {}Diagonal Las Torres 2640, Peñalol\'en, Santiago, Chile}}
\maketitle

\begin{abstract}
By means of the Lie algebra expansion method, the centrally extended
conformal algebra in two dimensions and the $\mathfrak{bms}_{3}$ algebra are
obtained from the Virasoro algebra. We extend this result to construct new
families of expanded Virasoro algebras that turn out to be infinite
dimensional lifts of the so-called $\mathfrak{B}_{k}$, $\mathfrak{C}_{k}$
and $\mathfrak{D}_{k}$ algebras recently introduced in the literature in the
context of (super)gravity. We also show how some of these new
infinite-dimensional symmetries can be obtained from expanded Ka\v{c}-Moody
algebras using modified Sugawara constructions. Applications in the context
of three-dimensional gravity are briefly discussed.
\end{abstract}

\bigskip

\newpage

\section{Introduction}

Infinite dimensional symmetries play a prominent role in different areas of
physics. In particular, symmetries of the Virasoro type have had remarkable
applications in two-dimensional field theory, fluid mechanics, string
theory, soliton theory and gravity among others.

The Virasoro algebra corresponds to the central extension of the algebra of
infinitesimal diffeomorphisms of the circle \cite{Goddard1988}. It was first
found in the context of string theory, where it describes the conformal
invariance of the two-dimensional worldsheet swept out by strings. This is
due to the fact that the conformal algebra in two dimensions is infinite
dimensional and its central extension is given by two copies the Virasoro
algebra. Therefore, the Virasoro symmetry appears in any physical system
with conformal invariance defined on a two-dimensional space. Examples of
this are two-dimensional sigma models \cite{Jimbo2012}, spin lattice models
near criticality \cite{Ketov2013}, integrable systems of the KdV type \cite%
{Guha2005}, the asymptotic structure of the $\mathcal{S}$-matrix in General
Relativity \cite{Kapec2014} and the asymptotic symmetries of
three-dimensional gravity. In the last case, Brown and Henneaux \cite{BH}
showed that for suitable boundary conditions the asymptotic symmetry of
three-dimensional Einstein gravity with negative cosmological constant is
given by two copies of the Virasoro algebra. The presence of the centrally
extended 2D-conformal symmetry at infinity was the first hint of an
holographic duality, which was later conjectured by Maldacena in the context
of strings \cite{Maldacena1995}. This remarkable discovery has lead to a
number of important subsequent results that could shed light into the
quantum nature of gravity \cite{Strominger1996, Strominger1998}.

The Virasoro algebra is closely related to the Ka\v{c}-Moody algebra, which
corresponds to the central extensions of the loop algebra. In fact, a
representation of the Virasoro algebra can be constructed out of quadratic
combinations of the generators of the Ka\v{c}-Moody algebra by means of the
Sugawara construction. This is useful when studying WZW models, as it allows
one to find the Virasoro symmetry at the level of the energy momentum tensor
starting from its current algebra \cite{DiFrancesco}. Furthermore the
Drinfeld-Sokolov Hamiltonian reduction relates the WZW model to Liouville
theory, which is conformally invariant \cite{BO}. In the context of
three-dimensional gravity this has had remarkable uses. In fact, when 3D
Einstein gravity with negative cosmological constant is formulated as a
Chern-Simons theory, it can be written as an $SL(2,\mathbb{R})$ WZW model
once the Hamiltonian constraints are solved within the action. Then, upon
imposing the Brown-Henneaux boundary conditions, it can be further reduced
to Liouville theory at the boundary \cite{CHV}.

The Virasoro symmetry is not the only infinite dimensional symmetry that
appears when studying the asymptotic structure of gravity theories. In
four-dimensional General Relativity, the BMS group appears as the asymptotic
symmetry of the theory at null infinity \cite{BBM, Sachs1, Sachs2}. This
symmetry has attracted great attention in the last years regarding the $%
\mathcal{S}$-Matrix for quantum gravity and its connection with soft
theorems and the gravitational memory effect \cite{Strominger2016}.
Remarkably, this enhancement of the Poincaré symmetry can also be found in
three dimensions \cite{ABS}. Indeed, in the case of vanishing cosmological
constant, the $\mathfrak{bms}_{3}$ algebra is found as the asymptotic
symmetry of Einstein gravity at null infinity \cite{BT}. This algebra is
given by the semi-direct sum of the infinitesimal diffeomorphisms on the
circle with an abelian ideal of super translations and can be obtained as an
Inönü-Wigner (IW) contraction \cite{IW,WW} of the centrally extended
conformal algebra in two dimensions in the very same way as the Poincaré
symmetry follows from the AdS$_{3}$ symmetry \cite{BC}. The $\mathfrak{bms}%
_{3}$ algebra can also be obtained from a Sugawara construction of the $%
\mathfrak{iso}(2,1)$ current algebra associated to the flat WZW model, which
in turn follows from an IW contraction of an $\mathfrak{sl}(2)$ Ka\v{c}%
-Moody algebra. Along the same lines, a Hamiltonian reduction of the flat
WZW model leads to the flat Liouville theory as the classical
two-dimensional dual for asymptotically flat 3D Einstein gravity, which is
BMS$_{3}$ invariant \cite{BG, BGG, BGG1}. On the other hand, there is an
equivalence between symmetries of ultra-relativistic theory and $\mathfrak{%
bms}_{3}$ which has been relevant in the extension of the AdS/CFT
correspondence \cite{FN1, FN2, FN3, FNR}. In the last years, generalizations
of the conformal and $\mathfrak{bms}_{3}$ algebras together with their Ka%
\v{c}-Moody cousins have appeared in the literature in the context of
three-dimensional supergravity and higher spin gravity \cite{Henneaux2000,
Campoleoni2010, Troessaert, GMPT, BDMT, MPTT, FMT1, FMT2, BDMT2, BJLMN, LM,
DR1, BCG, SA, BLN, BDR, FMT}. Analogously to the pure gravity cases, these
extensions turn out to be connected by IW contractions.

A particular characteristic of the IW contraction is that the starting and
resulting algebras have the same number of generators. A natural way to
generalize the IW contraction in order to obtain algebras of greater
dimension than the starting one is given by the Lie algebra expansion method
\cite{Hatsuda, A1, A2, A3, Sexp}. In particular, the $S$-expansion procedure
formulated in ref. \cite{Sexp} combines the structure constants of a given
Lie algebra with the inner product of an abelian semigroup and has given
rise to a number of interesting new symmetries that can be used to formulate
gravity theories coupled to matter \cite{GRCS, DFIMRSV, CPRS1, SS, CPRS2,
CPRS3, CDMR, CDIMR, Durka, CMR}. Such symmetries can be classified into
three families of algebras called $\mathfrak{B}_{k}$, $\mathfrak{C}_{k}$ and
$\mathfrak{D}_{k}$. $\mathfrak{B}_{k}$ algebras have been used to obtain
General Relativiy from Chern-Simons and Born-Infeld gravity theories in
diverse dimensions \cite{GRCS, CPRS1, CPRS2, CPRS3}. In particular, the $%
\mathfrak{B}_{3}$ and $\mathfrak{B}_{4}$ algebras correspond to the Poincaré
and Maxwell algebras \cite{DKGS, BGKL1, BGKL2, AKL, SSV, HR}. It is
important to note that $\mathfrak{B}_{k}$ symmetries can be obtained as IW
contraction of the $\mathfrak{C}_{k}$ algebras \cite{SS}. The $\mathfrak{C}%
_{k}$ family allows one to relate diverse (pure) Lovelock gravities to
Chern-Simons and Born-Infeld gravities \cite{CDIMR, CMR}. Alternatively, $%
\mathfrak{B}_{k}$ algebras can be obtained as an IW contraction of another
set of algebras called $\mathfrak{D}_{k}$ symmetries, which correspond to
direct sums of the form $\mathrm{AdS}\oplus \mathfrak{B}_{k-2}$ \cite{CDMR,
Durka}. Supersymmetric extensions of some of these expanded algebras have
been worked out in refs. {\cite{AILW, CR1, CR2, CRS, CFRS, CIRR, CFR, PR},
which can also be obtained through the $S$-expansion mechanism. } It is
therefore interesting to study what kind of infinite dimensional symmetries
can be obtained as S-expansions of known infinite dimensional algebras. In
this paper we put forward such study and present new families of infinite
dimensional algebras that can be obtained by applying the semigroup
expansion mechanism to the Virasoro algebra. We first show that the
centrally extended 2D-conformal algebra and the $\mathfrak{bms}_{3}$ algebra
can be obtained as a semigroup expansion of the Virasoro algebra. Then, by
using more general semigroups, we construct new families expanded Virasoro
algebras that we name \textit{generalized 2D-conformal algebras} and \textit{%
generalized $\mathfrak{bms}_{3}$ algebras}. We also show how these new
infinite dimensional symmetries can be related by IW contractions.
Interestingly these symmetries correspond to infinite dimensional
enhancements of the $\mathfrak{B}_{k}$ and $\mathfrak{C}_{k}$ algebras.
Additionally, we provide with an infinite dimensional lift of the so called.
$\mathfrak{D}_{k}$ algebras. Finally we study the Sugawara construction
connecting expanded Ka\v{c}-Moody algebras with our expanded Virasoro
algebras and present some explicit examples.

The paper is organized as follows: In Sect. 2 we present the general setup
to $S$-expand the Virasoro algebra and obtain the centrally extended
2D-conformal algebra as well as the $\mathfrak{bms}_3$ algebra particular
cases. In Sect. 3 we show how the expansion procedure can be used to
construct a deformed $\mathfrak{bms}_3$ algebra which corresponds to an
infinite dimensional lift of the Maxwell algebra. In the same way, an
infinite dimensional enhancement of the AdS-Lorentz algebra is constructed,
which is given by three copies of the Virasoro algebra and can be related to
the deformed $\mathfrak{bms}_3$ symmetry by an IW contraction. In Sect. 4
we introduce the generalized 2D-conformal algebras and generalized $%
\mathfrak{bms}_3$ algebras. Sect. 5 is devoted to the construction of the
infinite dimensional lifts of the $\mathfrak{D}_k$ algebras. In Sect. 6 we
present (modified) Sugawara construction that allows one to obtain expanded
Virasoro algebras from expanded Kac-Moody algebras in the simplest cases.
Finally, future applications of these results in the context of 3D gravity
theories and WZW models are discussed in Sect. 7.

\section{Centrally extended 2D-conformal algebra and $\mathfrak{bms}_{3}$
algebra as $S$-expansions}

\setcounter{equation}{0}

The $S$-expansion method \cite{Sexp} consists in combining the structure
constants of a Lie algebra $\mathfrak{g}$ with the elements of a semigroup $%
S $ to obtain a new Lie algebra $\mathfrak{G}=S\times \mathfrak{g}$. In this
section we show that the centrally extended 2D-conformal algebra and
the $\mathfrak{bms}_{3}$ algebra can be obtained explicitly as an $S$%
-expansion of the Virasoro algebra for suitable semigroups. For details
regarding the notation and the $S$-expansion procedure we refer the reader
to the original references \cite{Caroca2010a, Caroca2010b, Caroca2011, CKMN,
AMNT, ACCSP, ILPR, IKMN}.

\subsection{Expanding the Virasoro algebra}

The starting point of this construction is the Virasoro algebra $\mathfrak{%
vir}$,
\begin{equation}
\left[\ell_{m},\ell_{n}\right]=(m-n)\ell_{m+n}+\frac{c}{12}%
m(m^{2}-1)\delta_{m+n,0}\,,  \label{vir}
\end{equation}
together with a semigroup $S=\left \{ \lambda_{\alpha}\right \} $, whose
inner product is defined by a $2$-selector $K_{\alpha
\beta}^{\gamma}=K_{\beta \alpha}^{\gamma}$ such that
\begin{equation}
\lambda_{\alpha}\lambda_{\beta}=\lambda_{\beta}\lambda_{\alpha}=K_{\alpha%
\beta}^{\gamma}\lambda_{\gamma}\,.  \label{Kselector}
\end{equation}
We define an $S$-expanded Virasoro algebra as the direct product $\mathfrak{%
vir}_{\mathfrak{h}}=S\times \mathfrak{vir}$, where $\mathfrak{h}=S\times
\mathfrak{sl}(2)$ is the expansion of the $\mathfrak{sl}(2,\mathbb{R})$
subalgebra of \eqref{vir} generated by subset $\left \{
\ell_{-1},\ell_{0},\ell_{1} \right \} $ \footnote{%
This notation might seem awkward, but throughout our presentation it will
prove useful to label expanded Virasoro algebras by their corresponding
subalgebras $\mathfrak{h}$.}. The generators of $\mathfrak{vir}_{\mathfrak{h}%
}$ are given by
\begin{equation}
\ell_{(m,\alpha)}=\lambda_{\alpha}\ell_{m}  \label{Lexp}
\end{equation}
and satisfy the commutation relations
\begin{equation}
\left[\ell_{\left(m,\alpha \right)},\ell_{\left(n,\beta \right)}\right]%
=(m-n)K_{\alpha \beta}^{\gamma}\, \ell_{\left(m+n,\gamma \right)}+\frac{%
c_{\alpha \beta}}{12}m(m^{2}-1)\delta_{m+n,0}\,.  \label{virS}
\end{equation}
where $c_{\alpha \beta}$ denotes a set of central charges given by
\begin{equation}
c_{\alpha \beta}=c\,K_{\alpha \beta}^{\gamma}\lambda_{\gamma} \,.
\label{cexp}
\end{equation}
Note that the finite subalgebra ${\mathfrak{h}}$ of $\mathfrak{vir}_{%
\mathfrak{h}}$ is spanned by the subset $\left \{
\ell_{(-1,\alpha)},\ell_{(0,\alpha)},\ell_{(1,\alpha)} \right \} $.

\subsection{Centrally extended 2D-conformal algebra}

The centrally extended conformal algebra in two-dimensions is given by the
direct sum of a pair of Virasoro algebras $\mathfrak{vir}\oplus \mathfrak{vir%
} $, which we will simply denote as $\mathfrak{vir}^2$,
\begin{equation}
\begin{array}{lcl}
\left[\mathcal{L}_{m},\mathcal{L}_{n}\right] & = & \left(m-n\right)\mathcal{L%
}_{m+n}+\dfrac{c}{12}m(m^{2}-1)\delta_{m+n,0}\,, \\[6pt]
\left[\mathcal{\bar{L}}_{m},\mathcal{\bar{L}}_{n}\right] & = &
\left(m-n\right)\mathcal{\bar{L}}_{m+n}+\dfrac{\bar{c}}{12}%
m(m^{2}-1)\delta_{m+n,0}\,, \\[6pt]
\left[\mathcal{L}_{m},\mathcal{\bar{L}}_{n}\right] & = & 0\,.%
\end{array}
\label{vir2}
\end{equation}
This algebra can be obtained as a particular $S$-expansion of $\mathfrak{vir}
$. In fact, let us consider the (semi)group $\mathbb{Z}_{2}=\left \{
\lambda_{0},\lambda_{1}\right \} $, whose multiplication law is given by
\begin{equation}
\begin{tabular}{l|ll}
$\lambda_{1}$ & $\lambda_{1}$ & $\lambda_{0}$ \\
$\lambda_{0}$ & $\lambda_{0}$ & $\lambda_{1}$ \\ \hline
& $\lambda_{0}$ & $\lambda_{1}$%
\end{tabular}
\label{Z2}
\end{equation}
and from which the non-vanishing $2$-selectors \eqref{Kselector} can be read
off to be $K_{00}^{0}=K_{11}^{0}=K_{01}^{1}=K_{10}^{1}=1$. Denoting the
generators \eqref{Lexp} and the central charges \eqref{cexp} of the
corresponding expanded algebra as
\begin{equation}
\begin{array}{lcl}
\mathcal{J}_{m} & \equiv & \ell_{(m,0)}=\lambda_{0}\ell_{m}\, \, \, \,,\, \,
\, \,c_{1}\equiv c_{00}=c_{11}=\lambda_{0}c\,, \\[5pt]
\mathcal{P}_{m} & \equiv & \ell_{(m,1)}=\lambda_{1}\ell_{m}\, \, \, \,,\, \,
\, \,c_{2}\equiv c_{01}=\lambda_{1}c\,,%
\end{array}
\label{redefL}
\end{equation}
equation \eqref{virS} yields%
\begin{equation}
\begin{array}{lcl}
\left[\mathcal{J}_{m},\mathcal{J}_{n}\right] & = & \left(m-n\right)\mathcal{J%
}_{m+n}+\dfrac{c_{1}}{12}\left(m^{3}-m\right)\delta_{m+n,0}\,, \\[5pt]
\left[\mathcal{J}_{m},\mathcal{P}_{n}\right] & = & \left(m-n\right)\mathcal{P%
}_{m+n}+\dfrac{c_{2}}{12}\left(m^{3}-m\right)\delta_{m+n,0}\,, \\[5pt]
\left[\mathcal{P}_{m},\mathcal{P}_{n}\right] & = & \left(m-n\right)\mathcal{J%
}_{m+n}+\dfrac{c_{1}}{12}\left(m^{3}-m\right)\delta_{m+n,0}\,.%
\end{array}
\label{virZ2}
\end{equation}%
It is easy to see that \eqref{virZ2} is isomorphic to $\mathfrak{vir}^2$ by
making the following change of basis
\begin{equation}  \label{redefvir2}
\mathcal{L}_{m}=\frac{1}{2}\left(\mathcal{P}_{m}+\mathcal{J}_{m}\right)\, \,
\, \, \,,\, \, \, \, \mathcal{\bar{L}}_{-m}=\frac{1}{2}\left(\mathcal{P}_{m}-%
\mathcal{J}_{m}\right)\,,
\end{equation}
which leads to \eqref{vir2} with central charges $c=\frac{1}{2}(c_{2}+c_{1})$
and $\bar{c}=\frac{1}{2}(c_{2}-c_{1})$.

\subsection{$\mathfrak{bms}_{3}$ algebra}

Consider now the expansion of the Viraroso algebra \eqref{vir} using the
semigroup $S_{E}^{\left( 1\right) }=\left\{ \lambda _{0},\lambda
_{1},\lambda _{2}\right\} $, whose elements satisfy the multiplication law
\begin{equation}
\begin{tabular}{l|lll}
$\lambda _{2}$ & $\lambda _{2}$ & $\lambda _{2}$ & $\lambda _{2}$ \\
$\lambda _{1}$ & $\lambda _{1}$ & $\lambda _{2}$ & $\lambda _{2}$ \\
$\lambda _{0}$ & $\lambda _{0}$ & $\lambda _{1}$ & $\lambda _{2}$ \\ \hline
& $\lambda _{0}$ & $\lambda _{1}$ & $\lambda _{2}$%
\end{tabular}
\label{SE1}
\end{equation}%
and where $\lambda _{2}\equiv 0_{S}$ is the zero element of the semigroup
such that $0_{S}\lambda _{\alpha }=0_{S}$. The $0_{S}$-reduced $%
S_{E}^{\left( 1\right) }$-expanded algebra is obtained imposing $0_{S}\times
\ell _{\left( m,\alpha \right) }=0$. Defining the non-vanishing expanded
generators \eqref{Lexp} in the same way as in \eqref{redefL}, we get
\begin{equation}
\begin{array}{lcl}
\left[ \mathcal{J}_{m},\mathcal{J}_{n}\right] & = & \left( m-n\right)
\mathcal{J}_{m+n}+\dfrac{c_{1}}{12}\left( m^{3}-m\right) \delta _{m+n,0}\,,
\\[6pt]
\left[ \mathcal{J}_{m},\mathcal{P}_{n}\right] & = & \left( m-n\right)
\mathcal{P}_{m+n}+\dfrac{c_{2}}{12}\left( m^{3}-m\right) \delta _{m+n,0}\,,
\\[6pt]
\left[ \mathcal{P}_{m},\mathcal{P}_{n}\right] & = & 0\,,%
\end{array}
\label{cbms3}
\end{equation}%
which corresponds to the $\mathfrak{bms}_{3}$ algebra \cite{BT}.

Let us also recall that the $\mathfrak{bms}_{3}$ algebra can be obtained
from two copies of the Virasoro algebra as an IW contraction. Writing $%
\mathfrak{vir}^{2}$ in the form \eqref{virZ2} and rescaling its generators
as
\begin{equation}
\mathcal{J}_{m}\rightarrow \mathcal{J}_{m}\,,\,\,\,\mathcal{P}%
_{m}\rightarrow \sigma \mathcal{P}_{m}\,,\,\,\,c_{1}\rightarrow
c_{1}\,,\,\,\,c_{2}\rightarrow \sigma c_{2}\,,  \label{iwbms}
\end{equation}%
leads to \eqref{cbms3} in the limit $\sigma \rightarrow \infty $. A similar
approach is considered in \cite{KRR} where they implemented the IW
contraction using a Grassman parameter. As we will see in the following,
this kind of limit procedure will be useful to establish different links
between more general expanded Virasoro algebras.

\section{Deformed $\mathfrak{bms}_{3}$ algebra}

\setcounter{equation}{0}

The centrally extended conformal algebra and its flat limit, the $\mathfrak{%
bms}_{3}$ algebra, are not the only symmetries that can be obtained using
the expansion method. In the present section we present new
infinite-dimensional symmetries which are directly obtained as an $S$%
-expansion of the Virasoro algebra. In particular, a deformed $\mathfrak{bms}%
_{3}$ algebra as well as three copies of the Virasoro algebra ($\mathfrak{vir%
}^{3}$) can be obtained, where the former corresponds to an IW contraction
of the latter.

\subsection{Deformed $\mathfrak{bms}_{3}$ as an $S$-expansion}

Let us consider the semigroup $S_{E}^{\left( 2\right) }=\left \{ \lambda
_{0},\lambda _{1},\lambda _{2},\lambda _{3}\right \} $, whose elements
satisfy
\begin{equation}
\begin{tabular}{l|llll}
$\lambda _{3}$ & $\lambda _{3}$ & $\lambda _{3}$ & $\lambda _{3}$ & $\lambda
_{3}$ \\
$\lambda _{2}$ & $\lambda _{2}$ & $\lambda _{3}$ & $\lambda _{3}$ & $\lambda
_{3}$ \\
$\lambda _{1}$ & $\lambda _{1}$ & $\lambda _{2}$ & $\lambda _{3}$ & $\lambda
_{3}$ \\
$\lambda _{0}$ & $\lambda _{0}$ & $\lambda _{1}$ & $\lambda _{2}$ & $\lambda
_{3}$ \\ \hline
& $\lambda _{0}$ & $\lambda _{1}$ & $\lambda _{2}$ & $\lambda _{3}$%
\end{tabular}
\label{sml}
\end{equation}%
and where $\lambda _{3}=0_{S}$ is the zero element. Denoting the generators %
\eqref{Lexp} and the central charges \eqref{cexp} of the corresponding
expanded algebra as
\begin{equation}
\begin{array}{lcl}
\mathcal{J}_{m} & \equiv & \ell _{(m,0)}=\lambda _{0}\ell _{m}\, \, \, \,,\,
\, \, \,c_{1}\equiv c_{00}=c_{11}=\lambda _{0}c\,, \\[5pt]
\mathcal{P}_{m} & \equiv & \ell _{(m,1)}=\lambda _{1}\ell _{m}\, \, \, \,,\,
\, \, \,c_{2}\equiv c_{01}=\lambda _{1}c\,, \\[5pt]
\mathcal{Z}_{m} & \equiv & \ell _{(m,2)}=\lambda _{2}\ell _{m}\, \, \, \,,\,
\, \, \,c_{3}\equiv c_{02}=c_{11}=\lambda _{2}c\,,%
\end{array}
\label{redefL-1}
\end{equation}%
the $0_{S}$-reduced $S_{E}^{\left( 2\right) }$-expanded algebra satisfies
the commutation relations
\begin{equation}
\begin{array}{lcl}
\left[ \mathcal{J}_{m},\mathcal{J}_{n}\right] & = & \left( m-n\right)
\mathcal{J}_{m+n}+\dfrac{c_{1}}{12}\left( m^{3}-m\right) \delta _{m+n,0}\,,
\\[6pt]
\left[ \mathcal{J}_{m},\mathcal{P}_{n}\right] & = & \left( m-n\right)
\mathcal{P}_{m+n}+\dfrac{c_{2}}{12}\left( m^{3}-m\right) \delta _{m+n,0}\,,
\\[6pt]
\left[ \mathcal{P}_{m},\mathcal{P}_{n}\right] & = & \left( m-n\right)
\mathcal{Z}_{m+n}+\dfrac{c_{3}}{12}\left( m^{3}-m\right) \delta _{m+n,0}\,,
\\[6pt]
\left[ \mathcal{J}_{m},\mathcal{Z}_{n}\right] & = & \left( m-n\right)
\mathcal{Z}_{m+n}+\dfrac{c_{3}}{12}\left( m^{3}-m\right) \delta _{m+n,0}\,,
\\[6pt]
\left[ \mathcal{P}_{m},\mathcal{Z}_{n}\right] & = & 0\, \\[6pt]
\left[ \mathcal{Z}_{m},\mathcal{Z}_{n}\right] & = & 0\,.%
\end{array}
\label{defbms}
\end{equation}%
Interestingly, the Maxwell algebra in $(2+1)$ dimensions is spanned by the
generators $\mathcal{J}_{0},\mathcal{J}_{1},\mathcal{J}_{-1},$ $\mathcal{P}%
_{0},\mathcal{P}_{1},\mathcal{P}_{-1}$ and $\mathcal{Z}_{0},\mathcal{Z}_{1},%
\mathcal{Z}_{-1}$. This can be made explicit in terms of generators $%
\left
\{ J_{a},P_{a},Z_{a}\right \} $ obtained through the following change
of basis \footnote{%
In this case the Maxwell algebra is realized with a non-diagonal Minkowski
metric $\eta _{ab}=%
\begin{pmatrix}
0 & 1 & 0 \\
1 & 0 & 0 \\
0 & 0 & 1%
\end{pmatrix}%
$}

\begin{equation}  \label{cv}
\begin{array}{lcl}
\mathcal{J}_{-1} & = & -2J_{0}\,,\text{ \ \ }\mathcal{J}_{0}=J_{2}\text{%
\thinspace },\text{ \ \ }\mathcal{J}_{1}=J_{1}\,, \\[6pt]
\mathcal{P}_{-1} & = & -2P_{0}\,,\text{ \ \ }\mathcal{P}_{0}=P_{2}\text{%
\thinspace },\text{ \ \ }\mathcal{P}_{1}=P_{1}\,, \\[6pt]
\mathcal{Z}_{-1} & = & -2Z_{0}\,,\text{ \ \ }\mathcal{Z}_{0}=Z_{2}\text{%
\thinspace },\text{ \ \ }\mathcal{Z}_{1}=Z_{1}\,.%
\end{array}%
\end{equation}
This means that the deformed $\mathfrak{bms}_3$ algebra \eqref{defbms}
corresponds to an infinite-dimensional lift of the $(2+1)$-dimensional
Maxwell algebra in the very same way as the algebras $\mathfrak{vir}^2$ and $%
\mathfrak{bms}_3$ are infinite-dimensional lifts of the $AdS$ and the
Poincar\'e algebras in $2+1$ dimensions respectively.

\subsection{Deformed $\mathfrak{bms}_{3}$ algebra as a limit of $\mathfrak{%
vir}^{3}$}

Let us consider now $S_{\mathcal{M}}^{\left(2\right)}=\left \{
\lambda_{0},\lambda_{1},\lambda_{2}\right \} $ as the relevant abelian
semigroup, whose elements satisfy the following multiplication law
\begin{equation}
\begin{tabular}{l|lll}
$\lambda_{2}$ & $\lambda_{2}$ & $\lambda_{1}$ & $\lambda_{2}$ \\
$\lambda_{1}$ & $\lambda_{1}$ & $\lambda_{2}$ & $\lambda_{1}$ \\
$\lambda_{0}$ & $\lambda_{0}$ & $\lambda_{1}$ & $\lambda_{2}$ \\ \hline
& $\lambda_{0}$ & $\lambda_{1}$ & $\lambda_{2}$%
\end{tabular}
\label{sml2}
\end{equation}
Unlike the $S_{E}^{\left(2\right)}$ semigroup, there is no zero element in
this case. Adopting the same notation (\ref{redefL-1}) for the generators of
the $S_{\mathcal{M}}^{\left(2\right)}$-expanded algebra, we find the
following commutation relations
\begin{equation}
\begin{array}{lcl}
\label{virr3} \left[\mathcal{J}_{m},\mathcal{J}_{n}\right] & = &
\left(m-n\right)\mathcal{J}_{m+n}+\dfrac{c_{1}}{12}\left(m^{3}-m\right)%
\delta_{m+n,0}\,, \\[6pt]
\left[\mathcal{J}_{m},\mathcal{P}_{n}\right] & = & \left(m-n\right)\mathcal{P%
}_{m+n}+\dfrac{c_{2}}{12}\left(m^{3}-m\right)\delta_{m+n,0}\,, \\[6pt]
\left[\mathcal{P}_{m},\mathcal{P}_{n}\right] & = & \left(m-n\right)\mathcal{Z%
}_{m+n}+\dfrac{c_{3}}{12}\left(m^{3}-m\right)\delta_{m+n,0}\,, \\[6pt]
\left[\mathcal{J}_{m},\mathcal{Z}_{n}\right] & = & \left(m-n\right)\mathcal{Z%
}_{m+n}+\dfrac{c_{3}}{12}\left(m^{3}-m\right)\delta_{m+n,0}\,,%
\end{array}%
\end{equation}

\begin{equation*}
\begin{array}{lcl}
\left[\mathcal{Z}_{m},\mathcal{Z}_{n}\right] & = & \left(m-n\right)\mathcal{Z%
}_{m+n}+\dfrac{c_{3}}{12}\left(m^{3}-m\right)\delta_{m+n,0}\,, \\[6pt]
\left[\mathcal{Z}_{m},\mathcal{P}_{n}\right] & = & \left(m-n\right)\mathcal{P%
}_{m+n}+\dfrac{c_{2}}{12}\left(m^{3}-m\right)\delta_{m+n,0}\,.%
\end{array}%
\end{equation*}
Note that the $AdS$-Lorentz algebra in $2+1$ dimensions, also known as the
semi-simple extension of the Poincar\'e algebra \cite{Sorokas}, is the
subalgebra of \eqref{virr3} spanned by the generators $\mathcal{J}_{0},%
\mathcal{J}_{1},\mathcal{J}_{-1}$, $\mathcal{P}_{0},\mathcal{P}_{1},\mathcal{%
P}_{-1}$ and $\mathcal{Z}_{0},\mathcal{Z}_{1},\mathcal{Z}_{-1}$. This can be
explicitly seen using the change of basis \eqref{cv}, showing that %
\eqref{virr3} defines and infinite dimensional lift of the $AdS$-Lorentz
algebra in $2+1$ dimensions.

Remarkably, there is a redefinition of the generators of \eqref{virr3} that
allows to see its true algebraic structure. In fact, considering the change
of basis
\begin{equation}  \label{redefvir3}
\mathcal{L}_{m} = \dfrac{1}{2}\left(\mathcal{P}_{m}+\mathcal{Z}%
_{m}\right)\,,\,\,\,\, \mathcal{\bar{L}}_{-m} = \dfrac{1}{2}\left(\mathcal{P}%
_{m}-\mathcal{Z}_{m}\right)\,,\,\,\,\, \mathcal{\tilde{L}}_{-m} = \mathcal{J}%
_{m}-\mathcal{Z}_{m}\,,
\end{equation}%
three copies of the Virasoro algebra, which will be denoted as $\mathfrak{vir%
}^3$, are revealed
\begin{equation}  \label{vir3}
\begin{array}{lcl}
\left[\mathcal{L}_{m},\mathcal{L}_{n}\right] & = & \left(m-n\right)\mathcal{L%
}_{m+n}+\dfrac{c}{12}\left(m^{3}-m\right)\delta_{m+n,0}\,, \\[5pt]
\left[\mathcal{\bar{L}}_{m},\mathcal{\bar{L}}_{n}\right] & = &
\left(m-n\right)\mathcal{\bar{L}}_{m+n}+\dfrac{\bar{c}}{12}%
\left(m^{3}-m\right)\delta_{m+n,0}\,, \\[5pt]
\left[\mathcal{\tilde{L}}_{m},\mathcal{\tilde{L}}_{n}\right] & = &
\left(m-n\right)\mathcal{\tilde{L}}_{m+n}+\dfrac{\tilde{c}}{12}%
\left(m^{3}-m\right)\delta_{m+n,0}\,.%
\end{array}%
\end{equation}
where the central extensions are given by $c=\frac{1}{2}\left(c_{2}+c_{3}%
\right)$, $\bar{c}=\frac{1}{2}\left(c_{2}-c_{3}\right)$ and $\tilde{c}%
=c_{1}-c_{3}$. Additionally, as in the case of the $\mathfrak{bms}_{3}$ and
the 2D-conformal algebra, there is a limit procedure relating $\mathfrak{vir}%
^3$ and the deformed $\mathfrak{bms}_{3}$ algebra through an IW contraction.
In fact, the following rescaling of the generators of \eqref{virr3},
\begin{equation}  \label{iwdefbms}
\begin{array}{lcl}
\mathcal{J}_{m} & \rightarrow & \mathcal{J}_{m}\,,\,\,\,\mathcal{P}%
_{m}\,\,\,\rightarrow\,\,\, \sigma \mathcal{P}_{m}\,,\,\,\,\mathcal{Z}%
_{m}\,\,\,\rightarrow\,\,\, \sigma^{2}\mathcal{Z}_{m}\,, \\
c_{1} & \rightarrow & c_{1}\,,\,\,\,\,\,\,\,\,\,c_{2}\,\,\,\,\rightarrow
\,\,\,\sigma c_{2}\,,\,\,\,\,\,\,\,\,c_{3}\,\,\,\,\rightarrow
\,\,\,\sigma^{2}c_{3}\text{ },%
\end{array}%
\end{equation}
leads to the deformed $\mathfrak{bms}_3$ algebra \eqref{defbms} in the limit
$\sigma \rightarrow \infty$.

\section{Generalized expanded Virasoro algebras}

\numberwithin{equation}{section}

In the previous sections we have seen how the $S$ expansion mechanism allows
one to obtain the centrally extended 2D-conformal algebra and the $\mathfrak{%
bms}_3$ algebra from the Virasoro algebra. In the context of
three-dimensional gravity, the centrally extended 2D-conformal algebra and
the $\mathfrak{bms}_3$ algebra correspond to infinite dimensional lifts of $%
AdS$ and the Poincar\'e symmetries in $2+1$ dimensions. Generalizing this
results we have subsequently shown how to construct infinite dimensional
lifts of the Maxwell and the $AdS$-Lorentz algebras in $2+1$ dimensions,
which correspond a deformed $\mathfrak{bms}_3$ symmetry in the former case
and to three copies of the Virasoro algebra in the latter. As it has
recently been pointed out in Refs. \citep{SS,CPRS2,CDMR,CR1}, the Poincar\'e
and the $AdS$ algebras as well as the Maxwell and the $AdS$-Lorentz algebras
correspond to particular cases of the $\mathfrak{B}_{k}$ and $\mathfrak{C}%
_{k}$ algebras for $k=3$ and $k=4$ respectively. Such families of algebras
have been of particular interest in the context of gravity. Indeed, as was
shown in Refs.~\citep{CPRS1,CPRS3,GRCS}, General Relativity can be obtained
as a particular limit of a Chern-Simons and a Born-Infeld gravity theory
using the $\mathfrak{B}_{k}$ symmetries. On the other hand, the $\mathfrak{C}%
_{k}$ algebras allow one to recover the pure Lovelock Lagrangian from
Chern-Simons and Born-Infeld theories \citep{CDIMR,CMR}.

The results obtained up to this point clearly suggest that, in the same way
as their respective finite subalgebras, the $\mathfrak{bms}_3$ and $%
\mathfrak{vir}^2$ algebras as well as the deformed $\mathfrak{bms}_3$ and
the $\mathfrak{vir}^3$ algebras should correspond to particular cases of
certain families of generalized infinite-dimensional symmetries. In this
section we present the general scheme that leads to such families of
expanded Virasoro algebras.

\subsection{Generalized $\mathfrak{bms}_{3}$ algebras}

Let $S_{E}^{\left( k-2\right) }=\left\{ \lambda _{0},\lambda _{1},\dots
,\lambda _{k-1}\right\} $ be the finite abelian semigroup whose elements
satisfy the following multiplication law
\begin{equation}
\lambda _{\alpha }\lambda _{\beta }=\left\{
\begin{array}{lcl}
\lambda _{\alpha +\beta }\,\,\, & \mathrm{if}\,\,\,\,\alpha +\beta \leq k-2
&  \\
\lambda _{k-1}\,\,\,\, & \mathrm{if}\,\,\,\,\,\,\alpha +\beta >k-2 &
\end{array}%
\right.  \label{sek2}
\end{equation}%
where $\lambda _{k-1}=0_{s}$ is the zero element of the semigroup. The $%
S_{E}^{\left( k-2\right) }$-expanded Virasoro algebra \eqref{virS} in this
case is given by
\begin{equation}
\left[ \ell _{\left( m,\alpha \right) },\ell _{\left( n,\beta \right) }%
\right] =\left\{
\begin{array}{lcl}
(m-n)\ell _{\left( m+n,\alpha +\beta \right) }+\dfrac{c_{\alpha +\beta +1}}{%
12}m(m^{2}-1)\delta _{m+n,0}\,\,\,\,\, & \mathrm{if}\,\,\,\,\alpha +\beta
\leq k-2 &  \\
0\,\,\,\, & \mathrm{if}\,\,\,\,\,\,\alpha +\beta >k-2 &
\end{array}%
\right.  \label{generalizedbms}
\end{equation}%
where we have defined $c_{\alpha +\beta +1}\equiv c_{\alpha \beta }$.
Following the notation introduced in section 2, the algebra %
\eqref{generalizedbms} will be denoted by $\mathfrak{vir}_{\mathfrak{B}_k}$,
as the subalgebra $\mathfrak{h}$ generated by $\left\{
\ell_{(-1,\alpha)},\ell_{(0,\alpha)},\ell_{(1,\alpha)}\right\}$ corresponds
to the the $\mathfrak{B}_{k}$ algebra in $2+1$ dimensions \cite{GRCS,
Zanelii}. It is easy to see that \eqref{sek2} always contains an abelian
ideal spanned by the subset of generators
\begin{equation*}
\mathcal{A}=\left\{ \ell _{\left( m,\tilde{\alpha}\right) }\right\}
\,\,\,\,,\,\,\,\,\text{$\tilde{\alpha}$}=\left[ \frac{k}{2}\right] ,\ldots
,k-2
\end{equation*}%
and for which
\begin{equation}
\begin{array}{lcl}
\left[ \ell _{\left( m,\tilde{\alpha}\right) },\ell _{\left( m,\tilde{\beta}%
\right) }\right] & = & 0\,, \\[5pt]
\left[ \ell _{\left( m,\tilde{\alpha}\right) },\ell _{\left( m,\alpha
\right) }\right] & \in & \mathcal{A}\,\,+\mathrm{central\,\,terms.} \\[5pt]
&  &
\end{array}
\label{virZ2-1}
\end{equation}%
For this reason the $\mathfrak{vir}_{\mathfrak{B}_k}$ algebra will be
referred to as \textit{generalized $\mathfrak{bms}_{3}$ algebra}. This
algebra corresponds to an infinite dimensional lift of the $\mathfrak{B}_{k}$
algebra in $2+1$ dimensions, which can be made explicit by redefining the
generators in the form
\begin{equation}
\begin{array}{lcl}
\mathcal{J}_{m}^{i} & \equiv & \ell _{(m,i)}=\lambda _{i}\ell _{m}\,, \\%
[5pt]
\mathcal{P}_{m}^{\bar{\imath}} & \equiv & \ell _{(m,\bar{\imath})}=\lambda _{%
\bar{\imath}}\ell _{m}\,,%
\end{array}
\label{redefLgen}
\end{equation}%
where $i$ takes even values and $\bar{\imath}$ takes odd values. Here we
identify the following cases:

\begin{itemize}
\item For $k-2=2N$ the abelian ideal $\mathcal{A}$ is generated by
\begin{equation*}
\mathcal{A}=\left \{
\begin{array}{lcl}
\mathcal{P}_{m}^{N+1},\mathcal{J}_{m}^{N+2},\ldots ,\mathcal{P}_{m}^{2N-1},%
\mathcal{J}_{m}^{2N}\, \, \, \, & \mathrm{for}\, \,N\, \text{\thinspace }%
\mathrm{even} &  \\
\mathcal{J}_{m}^{N+1},\mathcal{P}_{m}^{N+2},\ldots ,\mathcal{P}_{m}^{2N-1},%
\mathcal{J}_{m}^{2N}\, \, \, \, & \mathrm{for}\, \,N\, \text{\thinspace }%
\mathrm{odd} &
\end{array}%
\right.
\end{equation*}

\item For $k-2=2N+1$ the abelian ideal $\mathcal{A}$ is generated by
\begin{equation*}
\mathcal{A}=\left \{
\begin{array}{lcl}
\mathcal{P}_{m}^{N+1},\mathcal{J}_{m}^{N+2},\ldots ,\mathcal{J}_{m}^{2N},%
\mathcal{P}_{m}^{2N+1}\, \, \, \, & \mathrm{for}\, \,N\, \text{\thinspace }%
\mathrm{even} &  \\
\mathcal{J}_{m}^{N+1},\mathcal{P}_{m}^{N+2},\ldots ,\mathcal{J}_{m}^{2N},%
\mathcal{P}_{m}^{2N+1}\, \, \, \, & \mathrm{for}\, \,N\, \text{\thinspace }%
\mathrm{odd} &
\end{array}%
\right.
\end{equation*}
\end{itemize}

Using the definition \eqref{redefLgen}, one to write \eqref{generalizedbms}
in the form
\begin{equation}
\begin{array}{lcl}
\left[ \mathcal{J}_{m}^{i},\mathcal{J}_{n}^{j}\right] & = & \left(
m-n\right) \mathcal{J}_{m+n}^{i+j}+\dfrac{c_{i+j+1}}{12}\left(
m^{3}-m\right) \delta _{m+n,0}\,\,\,\,,\,\,\,\,\mathrm{for}\,\,\,\,i+j\leq
k-2\,, \\[5pt]
\left[ \mathcal{J}_{m}^{i},\mathcal{P}_{n}^{\bar{\imath}}\right] & = &
\left( m-n\right) \mathcal{P}_{m+n}^{i+\bar{\imath}}+\dfrac{c_{i+\bar{\imath}%
+1}}{12}\left( m^{3}-m\right) \delta _{m+n,0}\,\,\,\,,\,\,\,\,\mathrm{for}%
\,\,\,\,i+\bar{\imath}\leq k-2\,, \\[5pt]
\left[ \mathcal{P}_{m}^{\bar{\imath}},\mathcal{P}_{n}^{\bar{j}}\right] & = &
\left( m-n\right) \mathcal{J}_{m+n}^{\bar{\imath}+\bar{j}}+\dfrac{c_{\bar{%
\imath}+\bar{j}+1}}{12}\left( m^{3}-m\right) \delta _{m+n,0}\,\ ,\,\,\ \
\mathrm{for}\,\,\,\,\bar{\imath}+\bar{j}\leq k-2\,, \\
\mathrm{others} & = & 0\,.%
\end{array}
\label{virZ2-2}
\end{equation}%
As mentioned before, the $\mathfrak{B}_{k}$ algebra in $2+1$ dimensions is a
subalgebra of (\ref{virZ2-2}) spanned by the generators $\mathcal{J}_{0}^{i},%
\mathcal{J}_{1}^{i},\mathcal{J}_{-1}^{i}$ and $\mathcal{P}_{0}^{\bar{\imath}%
},\mathcal{P}_{1}^{\bar{\imath}},\mathcal{P}_{-1}^{\bar{\imath}}$.
Additionally, when written in this form it is straightforward to see that
setting $k=3$ leads to the $\mathfrak{bms}_{3}$ algebra \eqref{cbms3}, while
$k=4$ reproduces the deformed $\mathfrak{bms}_{3}$ algebra (\ref{defbms}).
Thus, $\mathfrak{bms}_{3}$ and its corresponding deformation can be
classified into the infinite family of generalized $\mathfrak{bms}_{3}$
algebras $\mathfrak{vir}_{\mathfrak{B}_k}$.

\subsection{Generalized 2D-conformal algebras}

\label{virCk}

Another family of expanded Virasoro algebras can be obtained by choosing a
different semigroup. Let us consider $S_{\mathcal{M}}^{\left( k-2\right)
}=\left\{ \lambda _{0},\lambda _{1},\dots ,\lambda _{k-2}\right\} $ as the
relevant abelian semigroup whose elements satisfy
\begin{equation}
\lambda _{\alpha }\lambda _{\beta }=\left\{
\begin{array}{lcl}
\lambda _{\alpha +\beta }\,\,\,\, & \mathrm{if}\,\,\,\,\alpha +\beta \leq k-2
&  \\
\lambda _{\alpha +\beta -2\left[ \frac{k-1}{2}\right] }\,\,\, & \mathrm{if}%
\,\,\,\,\alpha +\beta >k-2 &
\end{array}%
\right.  \label{smk2}
\end{equation}%
Then the $S_{\mathcal{M}}^{\left( k-2\right) }$-expanded algebra \eqref{virS}
takes the form
\begin{equation}
\left[ \ell _{\left( m,\alpha \right) },\ell _{\left( n,\beta \right) }%
\right] =\left\{
\begin{array}{lcl}
(m-n)\ell _{\left( m+n,\alpha +\beta \right) }+\dfrac{c_{\alpha +\beta +1}}{%
12}m(m^{2}-1)\delta _{m+n,0}\,\,\,\, & \mathrm{if}\,\,\,\,\alpha +\beta \leq
k-2 &  \\
(m-n)\ell _{\left( m+n,\alpha +\beta -2\left[ \frac{k-1}{2}\right] \right) }+%
\dfrac{c_{\alpha +\beta -2\left[ \frac{k-1}{2}\right] +1}}{12}%
m(m^{2}-1)\delta _{m+n,0}\,\,\,\, & \mathrm{if}\,\,\,\,\alpha +\beta >k-2 &
\end{array}%
\right.  \label{generalizedbms-1}
\end{equation}%
and corresponds to $\mathfrak{vir}_{\mathfrak{C}_{k}}$, as its subalgebra $%
\mathfrak{h}$ is given by the $\mathfrak{C}_{k}$ algebra in $2+1$ dimensions %
\citep{CDIMR,CMR}. This algebra corresponds to an infinite dimensional lift
of the $\mathfrak{C}_{k}$ algebra, which can be explicitly seen by
redefining the generators in the form (\ref{redefLgen}), yielding
\begin{equation}
\begin{array}{lcl}
\left[ \mathcal{J}_{m}^{i},\mathcal{J}_{n}^{j}\right] & = & \left(
m-n\right) \mathcal{J}_{m+n}^{\left\{ i+j\right\} }+\dfrac{c_{\left\{
i+j\right\} +1}}{12}\left( m^{3}-m\right) \delta _{m+n,0}\,, \\[5pt]
\left[ \mathcal{J}_{m}^{i},\mathcal{P}_{n}^{\bar{j}}\right] & = & \left(
m-n\right) \mathcal{P}_{m+n}^{\left\{ i+\bar{j}\right\} }+\dfrac{c_{\left\{
i+\bar{j}\right\} +1}}{12}\left( m^{3}-m\right) \delta _{m+n,0}\,, \\[5pt]
\left[ \mathcal{P}_{m}^{\bar{\imath}},\mathcal{P}_{n}^{\bar{j}}\right] & = &
\left( m-n\right) \mathcal{J}_{m+n}^{\left\{ \bar{\imath}+\bar{j}\right\} }+%
\dfrac{c_{\left\{ \bar{\imath}+\bar{j}\right\} +1}}{12}\left( m^{3}-m\right)
\delta _{m+n,0}\,,%
\end{array}
\label{virZ2-2-1}
\end{equation}%
where $\left\{ \cdots \right\} $ means%
\begin{equation}
\left\{ i+j\right\} =\left\{
\begin{array}{lcl}
i+j & \mathrm{if}\,\,\,\,i+j\leq k-2 &  \\
i+j-2\left[ \frac{k-1}{2}\right] \, & \mathrm{if}\,\,\,\,i+j>k-2 &
\end{array}%
\right.
\end{equation}%
As remarked before, the $\mathfrak{C}_{k}$ algebra in $2+1$ dimensions is
the subalgebra of $\mathfrak{vir}_{\mathfrak{C}_k}$ spanned by the
generators $\mathcal{J}_{0}^{i},\mathcal{J}_{1}^{i},\mathcal{J}_{-1}^{i}$
and $\mathcal{P}_{0}^{\bar{\imath}},\mathcal{P}_{1}^{\bar{\imath}},\mathcal{P%
}_{-1}^{\bar{\imath}}$. When written in the form \eqref{virZ2-2-1} it is
clear that setting $k=3$ leads to the centrally extended 2D-conformal
algebra \eqref{virZ2}, while the case $k=4$ leads to the $\mathfrak{vir}^3$
algebra \eqref{virr3}. Therefore $\mathfrak{vir}_{\mathfrak{C}_k}$ will be
referred to as (centrally extended) \textit{generalized 2D-conformal algebra}%
. As in the cases $k=3$ and $k=4$ studied in the previous sections, the
generalized 2D-conformal algebra can be related to the generalized $%
\mathfrak{bms}_{3}$ one through an IW contraction. In fact, rescaling the
generators of \eqref{virZ2-2-1} in the form
\begin{equation}  \label{iwgeneral}
\begin{array}{lcl}
\mathcal{J}_{m}^{i} & \rightarrow & \sigma ^{i}\mathcal{J}%
_{m}^{i}\,\,,\,\,\,\,\,\,\,\,\,\mathcal{P}_{m}^{\bar{j}}\,\,\,\rightarrow
\,\,\,\sigma ^{\bar{j}}\mathcal{P}_{m}^{\bar{j}}\,, \\
c_{i+1} & \rightarrow & \sigma ^{i}c_{i+1}\,,\,\,\,\,\,\,\,c_{\bar{\imath}%
+1}\,\,\,\rightarrow\,\,\, \sigma ^{\bar{\imath}}c_{\bar{\imath}+1}%
\end{array}%
\end{equation}
leads to the generalized $\mathfrak{bms}_{3}$ algebra \eqref{virZ2-2} in the
limit $\sigma \rightarrow \infty $.

The fact that the $\mathfrak{vir}_{\mathfrak{C}_{k}}$ reduces to two and
three copies of the Virasoro algebra in the cases $k=3$ and $k=4$,
respectively, might make one think that it could generally be written as $k-1$
copies of the Virasoro algebra. However this is not true. Let us consider,
for instance, the $\mathfrak{vir}_{\mathfrak{C}_{5}}$ algebra. Renaming its
generators as $\mathcal{J}_{m}^{0}\equiv \mathcal{J}_{m}$, $\mathcal{P}%
_{m}^{1}\equiv\mathcal{P}_{m}$, $\mathcal{J}_{m}^{2}\equiv\mathcal{Z}_{m}$
and $\mathcal{P}_{m}^{3}\equiv\mathcal{R}_{m}$, this algebra can be directly
read off from \eqref{virZ2-2-1} to be
\begin{equation}
\begin{array}{lcl}
\label{genvir}\left[ \mathcal{J}_{m},\mathcal{J}_{n}\right] & = & \left(
m-n\right) \mathcal{J}_{m+n}+\dfrac{c_{1}}{12}\left( m^{3}-m\right) \delta
_{m+n,0}\,, \\[6pt]
\left[ \mathcal{J}_{m},\mathcal{P}_{n}\right] & = & \left( m-n\right)
\mathcal{P}_{m+n}+\dfrac{c_{2}}{12}\left( m^{3}-m\right) \delta _{m+n,0}\,,
\\[6pt]
\left[ \mathcal{P}_{m},\mathcal{P}_{n}\right] & = & \left( m-n\right)
\mathcal{Z}_{m+n}+\dfrac{c_{3}}{12}\left( m^{3}-m\right) \delta _{m+n,0}\,,
\\[6pt]
\left[ \mathcal{J}_{m},\mathcal{Z}_{n}\right] & = & \left( m-n\right)
\mathcal{Z}_{m+n}+\dfrac{c_{3}}{12}\left( m^{3}-m\right) \delta _{m+n,0}\,,
\\[6pt]
\left[ \mathcal{Z}_{m},\mathcal{Z}_{n}\right] & = & \left( m-n\right)
\mathcal{J}_{m+n}+\dfrac{c_{1}}{12}\left( m^{3}-m\right) \delta _{m+n,0}\,,
\\[6pt]
\left[ \mathcal{Z}_{m},\mathcal{P}_{n}\right] & = & \left( m-n\right)
\mathcal{R}_{m+n}+\dfrac{c_{4}}{12}\left( m^{3}-m\right) \delta _{m+n,0}\,,
\\[6pt]
\left[ \mathcal{J}_{m},\mathcal{R}_{n}\right] & = & \left( m-n\right)
\mathcal{R}_{m+n}+\dfrac{c_{4}}{12}\left( m^{3}-m\right) \delta _{m+n,0}\,,
\\[6pt]
\left[ \mathcal{Z}_{m},\mathcal{R}_{n}\right] & = & \left( m-n\right)
\mathcal{P}_{m+n}+\dfrac{c_{2}}{12}\left( m^{3}-m\right) \delta _{m+n,0}\,,
\\[6pt]
\left[ \mathcal{R}_{m},\mathcal{R}_{n}\right] & = & \left( m-n\right)
\mathcal{Z}_{m+n}+\dfrac{c_{3}}{12}\left( m^{3}-m\right) \delta _{m+n,0}\,,
\\[6pt]
\left[ \mathcal{P}_{m},\mathcal{R}_{n}\right] & = & \left( m-n\right)
\mathcal{J}_{m+n}+\dfrac{c_{1}}{12}\left( m^{3}-m\right) \delta _{m+n,0}\,,
\\
&  &
\end{array}%
\end{equation}%
which cannot be redefined as four copies of the Virasoro algebra by means of
a generalization of \eqref{redefvir2} or \eqref{redefvir3}.

\section{Infinite dimensional $\mathfrak{D}_{k}$-like algebras}

\numberwithin{equation}{section}

In \cite{CDMR} new expanded algebras were presented as a family of
Maxwell-like algebras. Inspired by this construction, in this section we
consider the expansion of the Virasoro algebra by means of the semigroup $%
S_{D}^{\left( k-2\right) }$, defined by the product rule%
\begin{equation}
\lambda _{\alpha }\lambda _{\beta }=\left \{
\begin{array}{lcl}
\lambda _{\alpha +\beta }\, \, \, \, & \mathrm{if}\, \, \, \, \alpha +\beta
\leq k-2 &  \\
\lambda _{\left( \alpha +\beta -\left( k-1\right) \right) \mathrm{mod}2\text{
}+\left( k-3\right) } & \mathrm{if}\, \, \, \, \alpha +\beta >k-2 &
\end{array}%
\right.  \label{SemDk}
\end{equation}%
Using the notation (\ref{redefLgen}) for the expanded generators, the $%
S_{D}^{\left( k-2\right) }$-expanded algebra (\ref{virS}) can be written in
the form

\begin{equation}
\begin{array}{lcl}
\left[ \mathcal{J}_{m}^{i},\mathcal{J}_{n}^{j}\right] & = & \left(
m-n\right) \mathcal{J}_{m+n}^{\left\{ i+j\right\} }+\dfrac{c_{\left\{
i+j\right\} +1}}{12}\left( m^{3}-m\right) \delta _{m+n,0}\,, \\[5pt]
\left[ \mathcal{J}_{m}^{i},\mathcal{P}_{n}^{\bar{j}}\right] & = & \left(
m-n\right) \mathcal{P}_{m+n}^{\left\{ i+\bar{j}\right\} }+\dfrac{c_{\left\{
i+\bar{j}\right\} +1}}{12}\left( m^{3}-m\right) \delta _{m+n,0}\,, \\[5pt]
\left[ \mathcal{P}_{m}^{\bar{\imath}},\mathcal{P}_{n}^{\bar{j}}\right] & = &
\left( m-n\right) \mathcal{J}_{m+n}^{\left\{ \bar{\imath}+\bar{j}\right\} }+%
\dfrac{c_{\left\{ \bar{\imath}+\bar{j}\right\} +1}}{12}\left( m^{3}-m\right)
\delta _{m+n,0}\,,%
\end{array}
\label{virZ2-2-2}
\end{equation}%
where $\left\{ \cdots \right\} $ means
\begin{equation}
\left\{ i+j\right\} =\left\{
\begin{array}{lcl}
i+j & \mathrm{if}\,\,\,\,i+j\leq k-2 &  \\
(i+j-(k-1))\text{\textrm{mod}}2+(k-3)\, & \mathrm{if}\,\,\,\,i+j>k-2 &
\end{array}%
\right.
\end{equation}%
These algebra corresponds to $\mathfrak{vir}_{\mathfrak{D}_{k}}$, as their
subalgebra $\mathfrak{h}$ is given by the $\mathfrak{D}_{k}$ algebra in $2+1$
dimensions \cite{CDMR} and provides with an infinite dimensional lift of it.
Interestingly, this kind of algebras can be written as the direct sum of two
copies of the Virasoro algebra and a generalized $\mathfrak{bms}_{3}$
algebra, i.e, $\mathfrak{vir}_{\mathfrak{D}_{k}}=\mathfrak{vir}^{2}\oplus
\mathfrak{vir}_{\mathfrak{B}_{k-2}}$, when an appropriate change of basis is
considered. Furthermore, an IW contraction of $\mathfrak{vir}_{\mathfrak{D}%
_{k}}$ using the rescaling \eqref{iwgeneral} leads to the generalized $%
\mathfrak{bms}_3$ algebra $\mathfrak{vir}_{\mathfrak{B}_{k}}$. In the
following, a few simple examples will be worked out.

\subsection{$\mathfrak{vir}^{2}\oplus \mathfrak{bms}_{3}$}

The simplest case to consider\footnote{%
The semigroup \eqref{SemDk} is defined for $k>3$ and $k=4$ just gives the
semigroup $S_{\mathcal{M}}^{\left(2\right) }$, which was already studied in
section 3.} is $k=5$, for which \eqref{SemDk} yields the $\mathfrak{vir}_{%
\mathfrak{D}_{5}}$ algebra:
\begin{equation}
\begin{array}{lcl}
\left[ \mathcal{J}_{m},\mathcal{J}_{n}\right] & = & \left( m-n\right)
\mathcal{J}_{m+n}+\dfrac{c_{1}}{12}\left( m^{3}-m\right) \delta _{m+n,0}\,,
\\[6pt]
\left[ \mathcal{J}_{m},\mathcal{P}_{n}\right] & = & \left( m-n\right)
\mathcal{P}_{m+n}+\dfrac{c_{2}}{12}\left( m^{3}-m\right) \delta _{m+n,0}\,,
\\[6pt]
\left[ \mathcal{P}_{m},\mathcal{P}_{n}\right] & = & \left( m-n\right)
\mathcal{Z}_{m+n}+\dfrac{c_{3}}{12}\left( m^{3}-m\right) \delta _{m+n,0}\,,
\\[6pt]
\left[ \mathcal{J}_{m},\mathcal{Z}_{n}\right] & = & \left( m-n\right)
\mathcal{Z}_{m+n}+\frac{c_{3}}{12}\left( m^{3}-m\right) \delta _{m+n,0}\,, \\%
[6pt]
\left[ \mathcal{Z}_{m},\mathcal{Z}_{n}\right] & = & \left( m-n\right)
\mathcal{Z}_{m+n}+\dfrac{c_{3}}{12}\left( m^{3}-m\right) \delta _{m+n,0}\,,
\\[6pt]
\left[ \mathcal{Z}_{m},\mathcal{P}_{n}\right] & = & \left( m-n\right)
\mathcal{R}_{m+n}+\dfrac{c_{4}}{12}\left( m^{3}-m\right) \delta _{m+n,0}\,,
\\[6pt]
\left[ \mathcal{J}_{m},\mathcal{R}_{n}\right] & = & \left( m-n\right)
\mathcal{R}_{m+n}+\dfrac{c_{4}}{12}\left( m^{3}-m\right) \delta _{m+n,0}\,,
\\[6pt]
\left[ \mathcal{R}_{m},\mathcal{Z}_{n}\right] & = & \left( m-n\right)
\mathcal{R}_{m+n}+\dfrac{c_{4}}{12}\left( m^{3}-m\right) \delta _{m+n,0}\,,
\\[6pt]
\left[ \mathcal{R}_{m},\mathcal{P}_{n}\right] & = & \left( m-n\right)
\mathcal{Z}_{m+n}+\dfrac{c_{3}}{12}\left( m^{3}-m\right) \delta _{m+n,0}\,,
\\[6pt]
\left[ \mathcal{R}_{m},\mathcal{R}_{n}\right] & = & \left( m-n\right)
\mathcal{Z}_{m+n}+\dfrac{c_{3}}{12}\left( m^{3}-m\right) \delta _{m+n,0}\,,%
\end{array}
\label{vir2bms3}
\end{equation}%
where we have defined%
\begin{equation}
\mathcal{J}_{m}\equiv\mathcal{J}_{m}^{0}\,,\text{ \ \ }\mathcal{P}_{m}\equiv%
\mathcal{P}_{m}^{1}\text{\thinspace },\text{ \ \ }\mathcal{Z}_{m}\equiv%
\mathcal{J}_{m}^{2}\,,\text{ \ \ }\mathcal{R}_{m}\equiv\mathcal{P}_{m}^{3}\,.
\end{equation}%
The Maxwell-like algebra $\mathfrak{D}_{5}$ in $2+1$ dimensions \cite{CDMR}
is spanned by the generators $\mathcal{J}_{0},\mathcal{J}_{1},\mathcal{J}%
_{-1},$ $\mathcal{P}_{0},\mathcal{P}_{1},\mathcal{P}_{-1},$ $\mathcal{Z}_{0},%
\mathcal{Z}_{1},\mathcal{Z}_{-1}$ and $\mathcal{R}_{0},\mathcal{R}_{1},%
\mathcal{R}_{-1}$. The algebraic structure of the $\mathfrak{vir}_{\mathfrak{%
D}_{5}}$ algebra can be made manifest by performing a suitable change of
basis. Indeed, the following redefinition
\begin{equation}
\mathcal{L}_{m} =\frac{1}{2}\left( \mathcal{R}_{m}+\mathcal{Z}_{m}\right)
\,,\,\,\,\,\mathcal{\bar{L}}_{-m} = \frac{1}{2}\left( \mathcal{R}_{m}-%
\mathcal{Z}_{m}\right) \,,
\end{equation}%
reproduces the centrally extended 2D-conformal algebra \eqref{vir2} with
central charges $c=\frac{1}{2}\left( c_{4}+c_{3}\right)$ and $\bar{c}=\frac{1%
}{2}\left( c_{4}-c_{3}\right)$, while
\begin{equation}
\mathcal{\tilde{J}}_{m} =\frac{1}{2}\left( \mathcal{J}_{m}-\mathcal{Z}%
_{m}\right) \,,\,\,\,\,\mathcal{\tilde{P}}_{m} =\frac{1}{2}\left( \mathcal{P}%
_{m}-\mathcal{R}_{m}\right) \,,
\end{equation}
leads to the $\mathfrak{bms}_{3}$ algebra \eqref{cbms3} with central charges
$c_1=\frac{1}{2}\left( c_{1}-c_{3}\right)$ and $c_2=\frac{1}{2}\left(
c_{2}-c_{4}\right)$. Since the set of generators $\left \{ \mathcal{L}_{m},%
\mathcal{\tilde{L}}_{m}\right \} $ commutes with the set $\left \{ \mathcal{%
\tilde{J}}_{m},\mathcal{\tilde{P}}_{m}\right \} $, this shows that the $%
\mathfrak{vir}_{\mathfrak{D}_{5}}$ algebra is given by the direct sum of
these two subalgebras, namely, $\mathfrak{vir}^{2}\oplus \mathfrak{bms}_{3}$.

\subsection{$\mathfrak{vir}^{2}\oplus \,$deformed $\mathfrak{bms}_{3}$}

In the case $k=6$, the $S$-expanded algebra $\mathfrak{vir}_{\mathfrak{D}%
_{6}}$ is given by
\begin{equation}
\begin{array}{lcl}
\left[ \mathcal{J}_{m},\mathcal{J}_{n}\right]  & = & \left( m-n\right)
\mathcal{J}_{m+n}+\dfrac{c_{1}}{12}\left( m^{3}-m\right) \delta _{m+n,0}\,,
\\[6pt]
\left[ \mathcal{J}_{m},\mathcal{P}_{n}\right]  & = & \left( m-n\right)
\mathcal{P}_{m+n}+\dfrac{c_{2}}{12}\left( m^{3}-m\right) \delta _{m+n,0}\,,
\\[6pt]
\left[ \mathcal{P}_{m},\mathcal{P}_{n}\right]  & = & \left( m-n\right)
\mathcal{Z}_{m+n}+\dfrac{c_{3}}{12}\left( m^{3}-m\right) \delta _{m+n,0}\,,
\\[6pt]
\left[ \mathcal{J}_{m},\mathcal{Z}_{n}\right]  & = & \left( m-n\right)
\mathcal{Z}_{m+n}+\dfrac{c_{3}}{12}\left( m^{3}-m\right) \delta _{m+n,0}\,,
\\[6pt]
\left[ \mathcal{Z}_{m},\mathcal{Z}_{n}\right]  & = & \left( m-n\right)
\mathcal{M}_{m+n}+\dfrac{c_{5}}{12}\left( m^{3}-m\right) \delta _{m+n,0}\,,
\\[5pt]
\left[ \mathcal{Z}_{m},\mathcal{P}_{n}\right]  & = & \left( m-n\right)
\mathcal{R}_{m+n}+\dfrac{c_{4}}{12}\left( m^{3}-m\right) \delta _{m+n,0}\,,
\\[6pt]
\left[ \mathcal{J}_{m},\mathcal{R}_{n}\right]  & = & \left( m-n\right)
\mathcal{R}_{m+n}+\dfrac{c_{4}}{12}\left( m^{3}-m\right) \delta _{m+n,0}\,,
\\[6pt]
\left[ \mathcal{R}_{m},\mathcal{Z}_{n}\right]  & = & \left( m-n\right)
\mathcal{R}_{m+n}+\dfrac{c_{4}}{12}\left( m^{3}-m\right) \delta _{m+n,0}\,,
\\[6pt]
\left[ \mathcal{R}_{m},\mathcal{P}_{n}\right]  & = & \left( m-n\right)
\mathcal{M}_{m+n}+\dfrac{c_{5}}{12}\left( m^{3}-m\right) \delta _{m+n,0}\,,
\\[6pt]
\left[ \mathcal{R}_{m},\mathcal{R}_{n}\right]  & = & \left( m-n\right)
\mathcal{M}_{m+n}+\dfrac{c_{5}}{12}\left( m^{3}-m\right) \delta _{m+n,0}\,,
\\[6pt]
\left[ \mathcal{J}_{m},\mathcal{M}_{n}\right]  & = & \left( m-n\right)
\mathcal{M}_{m+n}+\dfrac{c_{5}}{12}\left( m^{3}-m\right) \delta _{m+n,0}\,,
\\[6pt]
\left[ \mathcal{M}_{m},\mathcal{P}_{n}\right]  & = & \left( m-n\right)
\mathcal{R}_{m+n}+\dfrac{c_{4}}{12}\left( m^{3}-m\right) \delta _{m+n,0}\,,
\\[6pt]
\left[ \mathcal{M}_{m},\mathcal{Z}_{n}\right]  & = & \left( m-n\right)
\mathcal{M}_{m+n}+\dfrac{c_{5}}{12}\left( m^{3}-m\right) \delta _{m+n,0}\,,
\\[6pt]
\left[ \mathcal{M}_{m},\mathcal{R}_{n}\right]  & = & \left( m-n\right)
\mathcal{R}_{m+n}+\dfrac{c_{4}}{12}\left( m^{3}-m\right) \delta _{m+n,0}\,,
\\[6pt]
\left[ \mathcal{M}_{m},\mathcal{M}_{n}\right]  & = & \left( m-n\right)
\mathcal{M}_{m+n}+\dfrac{c_{5}}{12}\left( m^{3}-m\right) \delta _{m+n,0}\,,%
\end{array}
\label{virD6}
\end{equation}%
where we have defined
\begin{equation}
\mathcal{J}_{m}\equiv \mathcal{J}_{m}^{0}\,,\text{ \ \ }\mathcal{P}%
_{m}\equiv \mathcal{P}_{m}^{1}\text{\thinspace },\text{ \ \ }\mathcal{Z}%
_{m}\equiv \mathcal{J}_{m}^{2}\,,\text{ \ \ }\mathcal{R}_{m}\equiv \mathcal{P%
}_{m}^{3},\text{ \ \ }\mathcal{M}_{m}\equiv \mathcal{J}_{m}^{4}\,.
\end{equation}%
The Maxwell-like algebra $\mathfrak{D}_{6}$ \cite{CDMR} is spanned by the
generators $\mathcal{J}_{0},\mathcal{J}_{1},\mathcal{J}_{-1},$ $\mathcal{P}%
_{0},\mathcal{P}_{1},\mathcal{P}_{-1},$ $\mathcal{Z}_{0},\mathcal{Z}_{1},%
\mathcal{Z}_{-1},$ $\mathcal{R}_{0},\mathcal{R}_{1},\mathcal{R}_{-1}$ and $%
\mathcal{M}_{0},\mathcal{M}_{1},\mathcal{M}_{-1}$. The algebraic structure
of this algebra can be unveiled by performing a suitable change of basis. In
fact, two copies of the Virasoro algebra with central charges $c=\frac{1}{2}%
\left( c_{4}+c_{5}\right) $ and $\bar{c}=\frac{1}{2}\left(
c_{4}-c_{5}\right) $ can be recovered considering the redefinition
\begin{equation}
\mathcal{L}_{m}=\frac{1}{2}\left( \mathcal{R}_{m}+\mathcal{M}_{m}\right)
\,,\,\,\,\,\mathcal{\bar{L}}_{-m}=\frac{1}{2}\left( \mathcal{R}_{m}-\mathcal{%
M}_{m}\right) \,.  \label{set1}
\end{equation}%
On the other hand, the change of basis
\begin{equation}
\mathcal{\check{J}}_{m}=\frac{1}{2}\left( \mathcal{J}_{m}-\mathcal{Z}%
_{m}\right) \,,\text{ \ \ }\mathcal{\check{P}}_{m}=\frac{1}{2}\left(
\mathcal{P}_{m}-\mathcal{R}_{m}\right) ,\text{ \ \ }\mathcal{\check{Z}}_{m}=%
\frac{1}{2}\left( \mathcal{Z}_{m}-\mathcal{M}_{m}\right) \,  \label{set2}
\end{equation}%
reproduces the deformed $\mathfrak{bms}_{3}$ algebra \eqref{defbms} with
central charges $c_{1}=\frac{1}{2}\left( c_{1}-c_{3}\right) $, $c_{2}=\frac{1%
}{2}\left( c_{2}-c_{4}\right) $ and $c_{3}=\frac{1}{2}\left(
c_{3}-c_{5}\right) $. Thus, as the generators \eqref{set1} commute with the
generators \eqref{set2}, the $S_{D}^{\left( 6\right) }$-expanded Virasoro
algebra $\mathfrak{vir}_{\mathfrak{D}_{6}}$ is isomorphic to $\mathfrak{vir}%
^{2}\oplus $\thinspace deformed $\mathfrak{bms}_{3}$. This procedure can be
generalized for higher values of $k$, showing that $\mathfrak{vir}_{%
\mathfrak{D}_{k}}=\mathfrak{vir}^{2}\oplus \mathfrak{vir}_{\mathfrak{B}%
_{k-2}}$ holds generically.


\section{Sugawara construction and expanded Virasoro algebras}

\numberwithin{equation}{section}

The Ka\v{c}-Moody algebra $\mathfrak{\hat{\mathfrak{g}}}_{k}$ corresponds to
the central extension of the loop algebra of a semi-simple Lie algebra $%
\mathfrak{g}$ and is given by
\begin{equation}
\left[j_{m}^{a},j_{n}^{b}\right]=if_{\quad
c}^{ab}j_{m+n}^{c}+kmg^{ab}\delta_{m,-n}\,,  \label{eq:km}
\end{equation}
where $f_{abc}=-f_{bac}$ correspond to the structure constants of $\mathfrak{%
\mathfrak{g}}$ and $k$ denotes its central extension. The Sugawara
construction allows one to construct a representation of the Virasoro
algebra out of bilinear combinations of the generators of the Ka\v{c}-Moody
algebra by defining
\begin{equation}
\mathcal{\ell}_{m}=\frac{1}{2(k+C_{\mathfrak{g}})}g_{ab}%
\sum_{n}:j_{n}^{a}j_{m-n}^{b}:\,,  \label{eq:Lm}
\end{equation}
where $C_{\mathfrak{g}}$ is the dual Coxeter number of $\mathfrak{\mathfrak{g%
}}$, $g_{ab}$ is the corresponding Killing-Cartan metric and normal ordering
:: is defined as
\begin{equation*}
\sum_{n}:A_{n}B_{m-n}:=\sum_{n\leq-1}A_{n}B_{m-n}+\sum_{n>-1}B_{m-n}A_{n}\,.
\end{equation*}
In fact, one can easily check that such definition implies that $\ell_{m}$
has conformal weight one, $\left[\mathcal{\ell}_{m},j_{n}^{a}\right]%
=-nj_{m+n}^{a}$, and satisfies the Virasoro algebra \eqref{vir} with central
charge
\begin{equation}
c=\frac{k\mathrm{dim}\mathfrak{g}}{k+C_{\mathfrak{g}}}\,.
\end{equation}
where $\mathrm{dim}\mathfrak{g=}g_{ab}g^{ab}$ is the dimension of $\mathfrak{%
g}$.

\subsection{Modified Sugawara construction}

The modified Sugawara construction consists in defining new Virasoro
generators
\begin{equation*}
\tilde{\ell}_{m}=\mathcal{\ell}_{m}+img_{ab}\alpha^{a}j_{m}^{b}+\frac{1}{2}%
k\alpha^{2}\delta_{m,0}\,,
\end{equation*}
where $\text{$\alpha\in\mathfrak{g}$}$. Provided the generators $\mathcal{%
\ell}_{m}$ satisfy $\mathfrak{vir}$ with central charge $c$, the modified
generators $\tilde{\ell}{}_{n}$ form a new representation of the Virasoro
algebra, i.e.
\begin{equation*}
\left[\tilde{\ell}{}_{m},\tilde{\ell}{}_{n}\right]=(m-n)\tilde{\ell}{}_{m+n}+%
\frac{\tilde{c}}{12}\left(m^{3}-m\right)\delta_{m,-n}\,,
\end{equation*}
with a shifted central charge given by
\begin{equation*}
\tilde{c}=c+12k\alpha^{2}\,.
\end{equation*}

In the following, we will show how the expanded Virasoro algebras presented
in the previous sections can be obtained from Ka\v{c}-Moody algebras
associated with the $\mathfrak{B}_{k}$ and $\mathfrak{C}_{k}$ algebras through
generalized (modified) Sugawara constructions.

\subsection{$\mathfrak{bms}_{3}$ and the Sugawara construction}

Let us consider the following Ka\v{c}-Moody like algebra with a semi-direct
product structure:
\begin{equation}
\begin{array}{lcl}
\left[ j_{m}^{a},j_{n}^{b}\right] & = & if_{\quad
c}^{ab}j_{m+n}^{c}+k_{1}mg^{ab}\delta _{m,-n}\,, \\[5pt]
\left[ j_{m}^{a},p_{n}^{b}\right] & = & if_{\quad
c}^{ab}p_{m+n}^{c}+k_{2}mg^{ab}\delta _{m,-n}\,, \\[5pt]
\left[ p_{m}^{a},p_{n}^{b}\right] & = & 0\,,%
\end{array}
\label{km}
\end{equation}%
which can be obtained from an $S$-expansion of (\ref{eq:km}) using the
semigroup $S_{E}^{\left( 1\right) }$ (see the appendix). Now we introduce the
following quadratic combinations of the generators $j_{m}^{a}$ and $%
p_{m}^{a} $,
\begin{eqnarray*}
\mathcal{P}_{m} &=&\frac{1}{2k_{2}}g_{ab}\sum_{n}:p_{n}^{a}p_{m-n}^{b}:\,, \\
\mathcal{J}_{m} &=&\frac{1}{2k_{2}}g_{ab}\sum_{n}:\left(
j_{n}^{a}p_{m-n}^{b}+p_{n}^{a}j_{m-n}^{b}\right) :-\frac{k_{1}+2C_{\mathfrak{%
g}}}{k_{2}}\mathcal{P}_{m}\,.
\end{eqnarray*}%
Using the affine current algebra (\ref{km}) it is easy to see that
\begin{equation}
\begin{array}{lcl}
\left[ \mathcal{J}_{m},j_{n}^{a}\right] & = & -nj_{m+n}^{a}\,, \\[5pt]
\left[ \mathcal{J}_{m},p_{n}^{a}\right] & = & -np_{m+n}^{a}\,, \\[5pt]
\left[ \mathcal{P}_{m},j_{n}^{a}\right] & = & -np_{m+n}^{a}\,, \\
\left[ \mathcal{P}_{m},p_{n}^{a}\right] & = & 0\,,%
\end{array}
\label{JjPp}
\end{equation}%
and that $\mathcal{J}_{m}$ and $\mathcal{P}_{m}$ satisfy the commutation
relations
\begin{equation}
\begin{array}{lcl}
\left[ \mathcal{J}_{m},\mathcal{J}_{n}\right] & = & (m-n)\mathcal{J}_{m+n}+%
\frac{\mathrm{dim}\mathfrak{g}}{6}\left( m^{3}-m\right) \delta _{m,-n}\,, \\%
[5pt]
\left[ \mathcal{J}_{m},\mathcal{P}_{n}\right] & = & (m-n)\mathcal{P}_{m+n}\,,
\\[5pt]
\left[ \mathcal{P}_{m},\mathcal{P}_{n}\right] & = & 0\,,%
\end{array}
\label{bms3JP}
\end{equation}%
which corresponds to the $\mathfrak{bms}_{3}$ algebra \eqref{cbms3} with
central charges $c_{1}=2\,\mathrm{dim}\mathfrak{g}$ and $c_{2}=0$. The
central charge $c_{1}$ is familiar from the study of abelian Ka\v{c}-Moody
algebras \cite{Goddard1988} and manifests here due to the abelian ideal
generated by $p_{m}^{a}$. Now we can use the modified Sugawara construction
to obtain the fully centrally extended $\mathfrak{bm3}_3$ algebra from %
\eqref{km}. Indeed defining new generators:
\begin{eqnarray*}
\mathcal{\widetilde{J}}_{m} &=&\mathcal{J}_{m}+img_{ab}\alpha ^{a}j_{m}^{b}+%
\frac{1}{2}k_{1}\alpha ^{2}\delta _{m,0} \\
\mathcal{\widetilde{P}}_{m} &=&\mathcal{P}_{m}+img_{ab}\alpha ^{a}p_{m}^{b}+%
\frac{1}{2}k_{2}\alpha ^{2}\delta _{m,0}
\end{eqnarray*}%
one can show that
\begin{eqnarray*}
\left[ \mathcal{\mathcal{\widetilde{J}}}_{m},\mathcal{\mathcal{\widetilde{J}}%
}_{n}\right] &=&(m-n)\mathcal{\mathcal{\widetilde{J}}}_{m+n}+\frac{\tilde{%
c_{1}} }{12}\left( m^{3}-m\right) \delta _{m,-n} \\
\left[ \mathcal{\mathcal{\mathcal{\widetilde{J}}}}_{m},\mathcal{\widetilde{P}%
}_{n}\right] &=&(m-n)\mathcal{\widetilde{P}}_{m+n}+\frac{\tilde{c_{2}} }{12}%
\left( m^{3}-m\right) \delta _{m,-n} \\
\left[ \mathcal{\widetilde{P}}_{m},\mathcal{\widetilde{P}}_{n}\right] &=&0
\end{eqnarray*}%
where
\begin{eqnarray*}
\tilde{c_{1}} &=&\mathrm{2dim}\mathfrak{g}+12k_{1}\alpha ^{2} \\
\tilde{c_{2}} &=&12k_{2}\alpha ^{2}
\end{eqnarray*}%
This result can be understood as the quantum version of the Sugawara
construction described in \cite{BG,BDMT2} where $\mathfrak{bms}_{3} $ is
realized as a Poisson algebra for the central charges of asymptotically flat
three-dimensional Einstein gravity.


\subsection{$\mathfrak{vir}^{2}$ algebra}

The $\mathfrak{bms}_{3}$ algebra can also be obtained from the Sugawara
construction associated with a $\mathbb{Z}_2$-expansion of the Ka\v{c}-Moody
algebra, after an IW contraction. In fact, using the semigroup $\mathbb{Z}%
_2=S_{\mathcal{M}}^{\left(1\right)}$ to expand (\ref{eq:km}) (see the appendix),
we get
\begin{equation}
\begin{array}{lcl}
\left[j_{m}^{a},j_{n}^{b}\right] & = & if_{\quad
c}^{ab}j_{m+n}^{c}+k_{1}mg^{ab}\delta_{m,-n}\,, \\[5pt]
\left[j_{m}^{a},p_{n}^{b}\right] & = & if_{\quad
c}^{ab}p_{m+n}^{c}+k_{2}mg^{ab}\delta_{m,-n}\,, \\[5pt]
\left[p_{m}^{a},p_{n}^{b}\right] & = & if_{\quad
c}^{ab}j_{m+n}^{c}+k_{1}mg^{ab}\delta_{m,-n}\,.%
\end{array}%
\end{equation}
Redefining the generators as $j_{m}^{a}=l_{m}^{a}+\bar{l}_{-m}^{a}$ and $%
p_{m}^{a}=l_{m}^{a}-\bar{l}_{-m}^{a}$, this algebra can be written as the
product of two identical Ka\v{c}-Moody algebras with levels $k=\frac{1}{2}%
(k_{1}+k_{2})$ and $\bar{k}=\frac{1}{2}(k_{1}-k_{2})$, i.e,
\begin{eqnarray*}
\left[l_{m}^{a},l_{n}^{b}\right] & = & if_{\quad
c}^{ab}l_{m+n}^{c}+kmg^{ab}\delta_{m,-n}\,, \\
\left[\bar{l}{}_{m}^{a},\bar{l}{}_{n}^{b}\right] & = & if_{\quad c}^{ab}\bar{%
l}{}_{m+n}^{c}+\bar{k}mg^{ab}\delta_{m,-n}\,, \\
\left[l{}_{m}^{a},\bar{l}{}_{n}^{b}\right] & = & 0\,.
\end{eqnarray*}
This means that, considering two independent Sugawara constructions
\begin{eqnarray*}
\mathcal{\ell}_{m} & = & \frac{1}{2(k+C_{\mathfrak{g}})}g_{ab}%
\sum_{n}:l_{n}^{a}l_{m-n}^{b}:\,, \\
\mathcal{\bar{\ell}}_{m} & = & \frac{1}{2(\bar{k}+C_{\mathfrak{g}})}%
g_{ab}\sum_{n}:\bar{l}{}_{n}^{a}\bar{l}{}_{m-n}^{b}:\,,
\end{eqnarray*}
one can trivially obtain the $\mathfrak{vir}^{2}$ algebra (\ref{vir2}) with
central charges $c=\frac{k\mathrm{dim}\mathfrak{g}}{k+C_{\mathfrak{g}}}$ and
$\bar{c}=\frac{\bar{k}\mathrm{dim}\mathfrak{g}}{\bar{k}+C_{\mathfrak{g}}}$.
Using \eqref{redefvir2}, one can define
\begin{equation}  \label{eq:sugbms3-1}
\begin{array}{lcl}
\mathcal{P}_{m} & = & \dfrac{\sigma}{2(k+C_{\mathfrak{g}})}%
g_{ab}\sum_{n}:\left(l_{n}^{a}l_{m-n}^{b}+\mu\bar{l}{}_{n}^{a}\bar{l}{}%
_{-m-n}^{b}\right):\,, \\[6pt]
\mathcal{J}_{m} & = & \dfrac{1}{2(k+C_{\mathfrak{g}})}g_{ab}\sum_{n}:%
\left(l_{n}^{a}l_{m-n}^{b}-\mu\bar{l}{}_{n}^{a}\bar{l}{}_{-m-n}^{b}\right):%
\,,%
\end{array}%
\end{equation}
where $\mu=\frac{k+C_{\mathfrak{g}}}{\bar{k}+C_{\mathfrak{g}}}$. The
bilinears \eqref{eq:sugbms3-1} satisfy the $\mathfrak{bms}_{3} $ algebra (%
\ref{cbms3}) in the limit $\sigma\rightarrow\infty$ with central charges $%
c_{1}=$$\frac{\left(k-\mu\bar{k}\right)}{k+C_{\mathfrak{g}}}\mathrm{dim}%
\mathfrak{g}$ and $c_{2}=$$\frac{\left(k+\mu\bar{k}\right)}{k+C_{\mathfrak{g}%
}}\mathrm{dim}\mathfrak{g}$.


\subsection{Deformed $\mathfrak{bms}_{3}$ algebra from a Sugawara
construction}

The Sugawara construction presented before can be generalized in order to
recover the deformed $\mathfrak{bms}_{3}$ algebra \eqref{defbms} from an
expanded Ka\v{c}-Moody algebra. In this case we introduce the following
deformed current algebra:
\begin{equation}  \label{defkm}
\begin{array}{lcl}
\left[j_{m}^{a},j_{n}^{b}\right] & = & if_{\quad
c}^{ab}j_{m+n}^{c}+k_{1}mg^{ab}\delta_{m,-n}\,, \\[5pt]
\left[j_{m}^{a},p_{n}^{b}\right] & = & if_{\quad
c}^{ab}p_{m+n}^{c}+k_{2}mg^{ab}\delta_{m,-n}\,, \\[5pt]
\left[j_{m}^{a},z_{n}^{b}\right] & = & if_{\quad
c}^{ab}z_{m+n}^{c}+k_{3}mg^{ab}\delta_{m,-n}\,, \\[5pt]
\left[p_{m}^{a},p_{n}^{b}\right] & = & if_{\quad
c}^{ab}z_{m+n}^{c}+k_{3}mg^{ab}\delta_{m,-n}\,, \\[5pt]
\left[p_{m}^{a},z{}_{n}^{b}\right] & = & 0\,, \\[5pt]
\left[z_{m}^{a},z{}_{n}^{b}\right] & = & 0\,,%
\end{array}%
\end{equation}
which corresponds to an $S$-expansion of the Ka\v{c}-Moody algebra \eqref{km}
with the semigroup $S_{E}^{(2)}$ given in \eqref{sml} (see the appendix). Now we
define the following quadratic combinations of its generators:
\begin{equation}  \label{JPZ}
\begin{array}{lcl}
\mathcal{Z}_{m} & = & \dfrac{1}{2k_{3}}g_{ab}\sum_{n}:z_{n}^{a}z_{m-n}^{b}:%
\,, \\
\mathcal{P}_{m} & = & \dfrac{1}{2k_{3}}g_{ab}\sum_{n}:%
\left(p_{n}^{a}z_{m-n}^{b}+z_{n}^{a}p_{m-n}^{b}\right):-\dfrac{k_{2}}{k_{3}}%
\mathcal{Z}_{m}\,, \\
\mathcal{J}_{m} & = & \dfrac{1}{2k_{3}}g_{ab}\sum_{n}:%
\left(p_{n}^{a}p_{m-n}^{b}+j_{n}^{a}z_{m-n}^{b}+z_{n}^{a}j_{m-n}^{b}\right):-%
\dfrac{k_{2}}{k_{3}}\mathcal{P}_{m}-\dfrac{k_{1}+3C_{\mathfrak{g}}}{k_{3}}%
\mathcal{Z}_{m}\,.%
\end{array}%
\end{equation}
The commutators of $\mathcal{J}_{m}$, $\mathcal{P}_{m}$ and $\mathcal{Z}_{m}$
with the generators of \eqref{defkm} read
\begin{equation}  \label{JPZjpz}
\begin{array}{lcl}
\left[\mathcal{J}_{m},j_{n}^{a}\right]=-nj_{m+n}^{a}\,,\ \  & \left[\mathcal{%
P}_{m},j_{n}^{a}\right]=-np_{m+n}^{a}\,,\ \  & \left[\mathcal{Z}%
_{m},j_{n}^{a}\right]=-nz_{m+n}^{a}\,, \\[5pt]
\left[\mathcal{J}_{m},p_{n}^{a}\right]=-np_{m+n}^{a}\,,\ \  & \left[\mathcal{%
P}_{m},p_{n}^{a}\right]=-nz_{m+n}^{a}\,,\ \  & \left[\mathcal{Z}%
_{m},p_{n}^{a}\right]=0\,, \\[5pt]
\left[\mathcal{J}_{m},z_{n}^{a}\right]=-nz_{m+n}^{a}\,,\ \  & \left[\mathcal{%
P}_{m},z_{n}^{a}\right]=0\,,\ \ \ \ \ \ \ \ \ \ \  & \left[\mathcal{Z}%
_{m},z_{n}^{a}\right]=0\,.%
\end{array}%
\end{equation}
Using these relations the algebra of the bilinears \eqref{JPZ} turns out to
be given by
\begin{equation}  \label{defbmsJPZ}
\begin{array}{lcl}
\left[\mathcal{J}_{m},\mathcal{J}_{n}\right] & = & (m-n)\mathcal{J}_{m+n}+%
\dfrac{\mathrm{3dim}\mathfrak{g}}{12}\left(m^{3}-m\right)\delta_{m,-n}\,, \\%
[5pt]
\left[\mathcal{J}_{m},\mathcal{P}_{n}\right] & = & (m-n)\mathcal{P}_{m+n}\,,
\\[5pt]
\left[\mathcal{J}_{m},\mathcal{Z}_{n}\right] & = & (m-n)\mathcal{Z}_{m+n}\,,
\\[5pt]
\left[\mathcal{P}_{m},\mathcal{P}_{n}\right] & = & (m-n)\mathcal{Z}_{m+n}\,,
\\[5pt]
\left[\mathcal{P}_{m},\mathcal{Z}_{n}\right] & = & 0\,, \\[5pt]
\left[\mathcal{Z}_{m},\mathcal{Z}_{n}\right] & = & 0\,,%
\end{array}%
\end{equation}
which corresponds to the deformed $\mathfrak{bms}_{3}$ algebra \eqref{defbms}
with central charges $c_{1}=3\,\mathrm{dim}\mathfrak{g}$, $c_{2}=c_{3}=0$.
In order to obtain the fully centrally extended deformed $\mathfrak{bms}_{3}$
algebra from the deformed affine current algebra \eqref{defkm}, we introduce
the following modified Sugawara construction
\begin{equation}  \label{modSugJPZ}
\begin{array}{lcl}
\mathcal{\widetilde{J}}_{m} & = & \mathcal{J}_{m}+img_{ab}%
\alpha^{a}j_{m}^{b}+\frac{1}{2}k_{1}\alpha^{2}\delta_{m,0}\,, \\[5pt]
\mathcal{\widetilde{P}}_{m} & = & \mathcal{P}_{m}+img_{ab}%
\alpha^{a}p_{m}^{b}+\frac{1}{2}k_{2}\alpha^{2}\delta_{m,0}\,, \\[5pt]
\mathcal{\widetilde{Z}}_{m} & = & \mathcal{Z}_{m}+img_{ab}%
\alpha^{a}z_{m}^{b}+\frac{1}{2}k_{3}\alpha^{2}\delta_{m,0}\,.%
\end{array}%
\end{equation}
These generators satisfy the deformed $\mathfrak{bms}_3$ algebra,
\begin{eqnarray*}
\left[\mathcal{\mathcal{\widetilde{J}}}_{m},\mathcal{\mathcal{\widetilde{J}}}%
_{n}\right] & = & (m-n)\mathcal{\mathcal{\widetilde{J}}}_{m+n}+\frac{\tilde{%
c_{1}} }{12}\left(m^{3}-m\right)\delta_{m,-n}\,, \\
\left[\mathcal{\mathcal{\mathcal{\widetilde{J}}}}_{m},\mathcal{\widetilde{P}}%
_{n}\right] & = & (m-n)\mathcal{\widetilde{P}}_{m+n}+\frac{\tilde{c_{2}} }{12%
}\left(m^{3}-m\right)\delta_{m,-n}\,, \\
\left[\mathcal{\widetilde{J}}_{m},\mathcal{\widetilde{Z}}_{n}\right] & = &
(m-n)\mathcal{\widetilde{Z}}_{m+n}+\frac{\tilde{c_{3}} }{12}%
\left(m^{3}-m\right)\delta_{m,-n} \,, \\
\left[\mathcal{\widetilde{P}}_{m},\mathcal{\widetilde{P}}_{n}\right] & = &
(m-n)\mathcal{\widetilde{Z}}_{m+n}+\frac{\tilde{c_{3}} }{12}%
\left(m^{3}-m\right)\delta_{m,-n}\,, \\
\left[\mathcal{\widetilde{P}}_{m},\mathcal{\widetilde{Z}}_{n}\right] & = &
0\,, \\
\left[\mathcal{\widetilde{Z}}_{m},\mathcal{\widetilde{Z}}_{n}\right] & = &
0\,.
\end{eqnarray*}
where the central charges are given by
\begin{eqnarray*}
\tilde{c_{1}} & = & \mathrm{3dim}\mathfrak{g}+12k_{1}\alpha^{2} \,, \\
\tilde{c_{2}} & = & 12k_{2}\alpha^{2}\,, \\
\tilde{c_{3}} & = & 12k_{3}\alpha^{2}\,.
\end{eqnarray*}

\subsection{$\mathfrak{vir}^{3}$ algebra}

The deformed $\mathfrak{bms}_{3}$ algebra can also be obtained as an IW
contraction of the Sugawara construction associated with an $S$-expansion of
the Ka\v{c}-Moody algebra using the semigroup $S_{\mathcal{M}%
}^{\left(2\right)}$. In fact the $S_{\mathcal{M}}^{\left(2\right)}$-expanded
Ka\v{c}-Moody algebra is given by (see the appendix)
\begin{equation}
\begin{array}{lcl}
\left[j_{m}^{a},j_{n}^{b}\right] & = & if_{\quad
c}^{ab}j_{m+n}^{c}+k_{1}mg^{ab}\delta_{m,-n}\,, \\[5pt]
\left[j_{m}^{a},p_{n}^{b}\right] & = & if_{\quad
c}^{ab}p_{m+n}^{c}+k_{2}mg^{ab}\delta_{m,-n}\,, \\[5pt]
\left[j_{m}^{a},z_{n}^{b}\right] & = & if_{\quad
c}^{ab}z_{m+n}^{c}+k_{3}mg^{ab}\delta_{m,-n}\,, \\[5pt]
\left[p_{m}^{a},p_{n}^{b}\right] & = & if_{\quad
c}^{ab}z_{m+n}^{c}+k_{3}mg^{ab}\delta_{m,-n}\,, \\[5pt]
\left[z{}_{m}^{a},p_{n}^{b}\right] & = & if_{\quad
c}^{ab}p_{m+n}^{c}+k_{2}mg^{ab}\delta_{m,-n}\,, \\[5pt]
\left[z_{m}^{a},z{}_{n}^{b}\right] & = & if_{\quad
c}^{ab}z_{m+n}^{c}+k_{3}mg^{ab}\delta_{m,-n}\,.%
\end{array}
\label{defkm-1}
\end{equation}
which, through the redefinitions $z_{m}^{a}=l_{m}^{a}+\bar{l}_{-m}^{a}$, $%
p_{m}^{a}=l_{m}^{a}-\bar{l}_{-m}^{a}$ and $j_{m}^{a}=\tilde{l}%
_{m}^{a}+l_{m}^{a}+\bar{l}_{-m}^{a}$, can be written as the direct product
of three identical commuting Ka\v{c}-Moody algebras with levels $k=\frac{%
k_{3}+k_{2}}{2}$, $\bar{k}=\frac{k_{3}-k_{2}}{2}$ and $\tilde{k}=k_{1}-k_{3}$%
:
\begin{eqnarray*}
\left[l_{m}^{a},l_{n}^{b}\right] & = & if_{\quad
c}^{ab}l_{m+n}^{c}+kmg^{ab}\delta_{m,-n}\,, \\
\left[\bar{l}{}_{m}^{a},\bar{l}{}_{n}^{b}\right] & = & if_{\quad c}^{ab}\bar{%
l}{}_{m+n}^{c}+\bar{k}mg^{ab}\delta_{m,-n}\,, \\
\left[\tilde{l}_{m}^{a},\tilde{l}_{n}^{b}\right] & = & if_{\quad c}^{ab}%
\tilde{l}_{m+n}^{c}+\tilde{k}mg^{ab}\delta_{m,-n}\,.
\end{eqnarray*}
Therefore, considering three independent Sugawara constructions
\begin{eqnarray*}
\mathcal{\ell}_{m} & = & \frac{1}{2(k+C_{\mathfrak{g}})}g_{ab}%
\sum_{n}:l_{n}^{a}l_{m-n}^{b}:\,, \\
\mathcal{\bar{\ell}}_{m} & = & \frac{1}{2(\bar{k}+C_{\mathfrak{g}})}%
g_{ab}\sum_{n}:\bar{l}{}_{n}^{a}\bar{l}{}_{m-n}^{b}:\,, \\
\tilde{\ell}_{m} & = & \frac{1}{2(\tilde{k}+C_{\mathfrak{g}})}g_{ab}\sum_{n}:%
\tilde{l}{}_{n}^{a}\tilde{l}{}_{m-n}^{b}:\,,
\end{eqnarray*}
one can trivially obtain the $\mathfrak{vir}^{3}$ algebra \eqref{vir3} with
central charges $c=\frac{k\mathrm{dim}\mathfrak{g}}{k+C_{\mathfrak{g}}}$ , $%
\bar{c}=\frac{\bar{k}\mathrm{dim}\mathfrak{g}}{\bar{k}+C_{\mathfrak{g}}}$
and $\tilde{c}=\frac{\tilde{k}\mathrm{dim}\mathfrak{g}}{\tilde{k}+C_{%
\mathfrak{g}}}$. This means that, using the relation \eqref{redefvir3}, one
can define
\begin{equation}  \label{eq:sugbms3-1-1}
\begin{array}{lcl}
\mathcal{Z}_{m} & = & \dfrac{\sigma^{2}}{2(k+C_{\mathfrak{g}})}%
g_{ab}\sum_{n}:\left(l_{n}^{a}l_{m-n}^{b}-\mu\bar{l}{}_{n}^{a}\bar{l}{}%
_{-m-n}^{b}\right):\,, \\[6pt]
\mathcal{P}_{m} & = & \dfrac{\sigma}{2(k+C_{\mathfrak{g}})}%
g_{ab}\sum_{n}:\left(l_{n}^{a}l_{m-n}^{b}+\mu\bar{l}{}_{n}^{a}\bar{l}{}%
_{-m-n}^{b}\right):\,, \\[6pt]
\mathcal{J}_{m} & = & \dfrac{1}{2(k+C_{\mathfrak{g}})}g_{ab}\sum_{n}:%
\left(l_{n}^{a}l_{m-n}^{b}-\mu\bar{l}_{n}^{a}\bar{l}_{-m-n}^{b}+\nu\tilde{l}%
_{n}^{a}\tilde{l}_{-m-n}^{b}\right):\,,%
\end{array}%
\end{equation}
where $\mu=\frac{k+C_{\mathfrak{g}}}{\bar{k}+C_{\mathfrak{g}}}$ and $\nu=%
\frac{k+C_{\mathfrak{g}}}{\tilde{k}+C_{\mathfrak{g}}}$. It is easy to verify
that the bilinear combinations \eqref{eq:sugbms3-1-1} satisfy the deformed $%
\mathfrak{bms}_{3}$ algebra \eqref{defbms} in the limit $\sigma\rightarrow%
\infty$ with central charges $c_{1}=$$\frac{\left(k-\mu\bar{k}\right)}{k+C_{%
\mathfrak{g}}}\mathrm{dim}\mathfrak{g}$, $c_{2}=$$\frac{\left(k+\mu\bar{k}%
\right)}{k+C_{\mathfrak{g}}}\mathrm{dim}\mathfrak{g}$ and $c_{3}=$$\frac{%
\left(k+\nu\tilde{k}\right)}{k+C_{\mathfrak{g}}}\mathrm{dim}\mathfrak{g}$.


\subsection{Generalization}

Following the same steps as described above, one can in principle always find a
generalized (modified) Sugawara construction that, given a semigroup $S,$
allows one to pass from the $S$-expanded Ka\v{c}-Moody algebra to the
corresponding $S$-expanded Virasoro algebra . As we have seen, the Sugawara
construction for the $\mathfrak{bms}_{3}$ algebra and for the deformed $%
\mathfrak{bms}_{3}$ algebra are quite cumbersome and therefore their
generalization for $\mathfrak{vir}_{\mathfrak{B}_{k}}$ with $k>4$ will not
be given here. In the case of the generalized conformal algebras $\mathfrak{%
vir}_{\mathfrak{C}_{k}}$, the Sugawara constructions presented here have
been somewhat straightforward, as the cases $k=3$ and $k=4$ correspond
the direct product of two and three copies of the Virasoro algebra,
respectively. However, as we have stressed in Sect. \ref{virCk}, for $k>4$
it is not true anymore that the $\mathfrak{vir}_{\mathfrak{C}_{k}}$
algebras can be written as products of single copies of the Virasoro algebra
and therefore the Sugawara construction will be more complicated.


\section{Comments and further developments}

In this paper we have presented the general setup to obtain new infinite
dimensional algebras by applying the $S$-expansion method to the Virasoro
algebra. Interestingly, the algebras obtained here contain known finite
algebras as subalgebras and inherit the way they are related between each
other. Indeed, the following diagram summarizes the IW contractions that
relate the Poincar\'e, $AdS$, Maxwell and $AdS$-Lorentz algebras in $2+1$
dimensions as well as their relation with the Lorentz algebra through
different $S$-expansions:
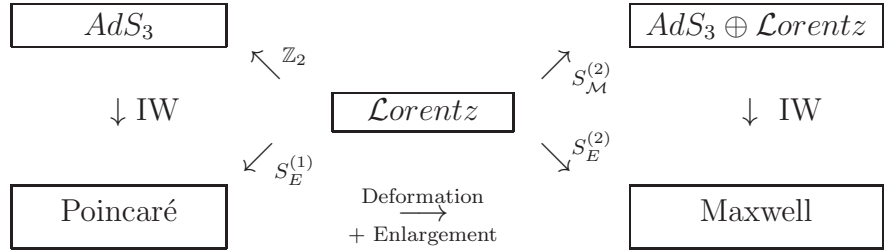
\begin{equation*}
\begin{tabular}{ccccc}
\cline{1-1}\cline{5-5}
\multicolumn{1}{|c}{$\overset{}{AdS_{3}}$} & \multicolumn{1}{|c}{} & $%
\overset{}{}$ &  & \multicolumn{1}{|c|}{$\overset{}{AdS_{3}\oplus \mathcal{L}%
orentz}$} \\ \cline{1-1}\cline{5-5}
& $\nwarrow ^{%
\mathbb{Z}
_{2}}$ &  & $\ \nearrow _{S_{\mathcal{M}}^{\left( 2\right) }}$ &  \\
\cline{3-3}
$\ \ \ \ \ \ \ \ \downarrow $ IW \ \ \  &  & \multicolumn{1}{|c}{$\mathcal{L}%
orentz$} & \multicolumn{1}{|c}{} & $\ \ \ \ \ \downarrow $ \ IW \\
\cline{3-3}
& $\swarrow _{S_{E}^{\left( 1\right) }}$ &  & $\ \searrow ^{S_{E}^{\left(
2\right) }}$ &  \\ \cline{1-1}\cline{5-5}
\multicolumn{1}{|c}{Poincaré} & \multicolumn{1}{|c}{} & $\underset{\text{+
Enlargement}}{\overset{\text{Deformation}}{\longrightarrow }}$ &  &
\multicolumn{1}{|c|}{Maxwell} \\ \cline{1-1}\cline{5-5}
\end{tabular}%
\end{equation*}%
In the first part of this article we have shown that the centrally extended
2D-conformal algebra $\mathfrak{vir}^{2}$ as well as the $\mathfrak{bms}_{3}$
algebra can be obtained as $S$-expansions of the Virasoro algebra using the
semigroups $\mathbb{Z}_{2}$ and $S_{E}^{(1)}$, respectively. Subsequently we
showed that, using the semigroups $S_{\mathcal{M}}^{(2)}$ and $S_{E}^{(2)}$,
the S-expansion leads to three copies of the Virasoro algebra in the former
case and to a deformed $\mathfrak{bms}_{3}$ algebra in the latter case. These
algebras correspond to infinite-dimensional lifts of the $AdS$-Lorentz and
Maxwell algebras and, furthermore, the deformed $\mathfrak{bms}_{3}$ algebra
can be obtained as an IW contraction of $\mathfrak{vir}^{3}$. This means
that the infinite dimensional symmetries presented here satisfy the same IW
contraction and expansion relations as their finite dimensional subalgebras
presented in the previous diagram, i.e.,
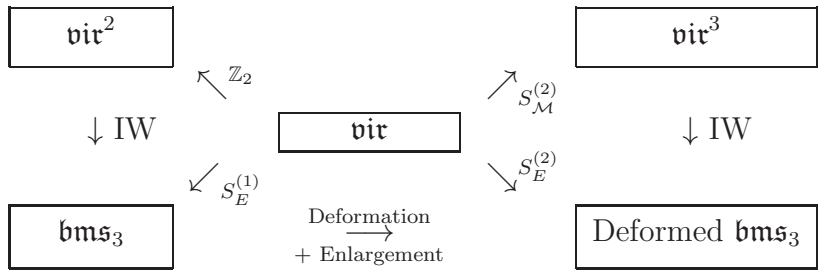
\begin{equation*}
\begin{tabular}{ccccc}
\cline{1-1}\cline{5-5}
\multicolumn{1}{|c}{$\overset{}{\mathfrak{vir}^{2}}$} & \multicolumn{1}{|c}{}
& $\underset{}{\overset{}{}}$ &  & \multicolumn{1}{|c|}{$\overset{}{%
\mathfrak{vir}^{3}}$} \\ \cline{1-1}\cline{5-5}
& $\nwarrow ^{%
\mathbb{Z}
_{2}}$ &  & $\ \nearrow _{S_{\mathcal{M}}^{\left( 2\right) }}$ &  \\
\cline{3-3}
\ \ \ \ \ \ $\downarrow $ IW &  & \multicolumn{1}{|c}{$\mathfrak{vir}$} &
\multicolumn{1}{|c}{} & $\ \ \ \ \downarrow $ IW \\ \cline{3-3}
& $\swarrow _{S_{E}^{\left( 1\right) }}$ &  & $\ \searrow ^{S_{E}^{\left(
2\right) }}$ &  \\ \cline{1-1}\cline{5-5}
\multicolumn{1}{|c}{$\overset{}{\mathfrak{bms}_{3}}$} & \multicolumn{1}{|c}{}
& $\underset{\text{+ Enlargement}}{\overset{\text{Deformation}}{%
\longrightarrow }}$ &  & \multicolumn{1}{|c|}{$\overset{}{\text{Deformed }%
\mathfrak{bms}_{3}}$} \\ \cline{1-1}\cline{5-5}
\end{tabular}%
\end{equation*}
In Sect. 4, we have generalized the previous results by considering
expansions of the Virasoro algebra with the semigroups $S^{(k-2)}_\mathcal{M}
$ and $S^{(k-2)}_E$ to obtain two sets of families of infinite-dimensional
algebras that we have called generalized $\mathfrak{bms}_3$ algebras and
generalized 2D-conformal algebras. These families are denoted, respectively,
by $\mathfrak{vir}_{\mathfrak{C}_k}$ and $\mathfrak{vir}_{\mathfrak{C}_k}$,
and reduce to the infinite-dimensional algebras previously discussed for $%
k=3 $ and $k=4$. Furthermore, they turn out to be related by an IW
contraction for every value of $k$
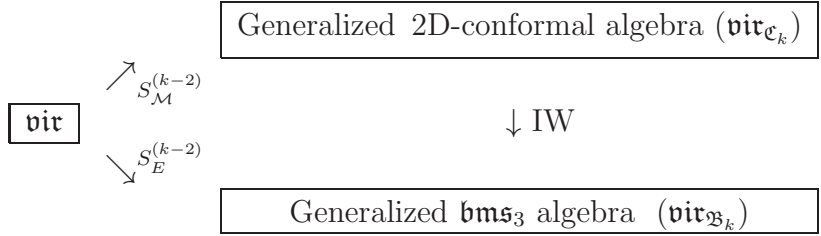
\begin{equation*}
\begin{tabular}{ccccc}
\cline{5-5}
&  & $\underset{}{\overset{}{}}$ &  & \multicolumn{1}{|c|}{$\overset{}{\text{%
Generalized\thinspace\ 2D-conformal}\,\,\text{algebra}\,\left(\mathfrak{vir}%
_{\mathfrak{C}_{k}}\right)}$} \\ \cline{5-5}
&  &  & $\ \nearrow _{S_{\mathcal{M}}^{\left( k-2\right) }}$ &  \\
\cline{3-3}
\ \ \ \ \ \  &  & \multicolumn{1}{|c}{$\mathfrak{vir}$} &
\multicolumn{1}{|c}{} & $\ \ \ \ \downarrow $ IW \\ \cline{3-3}
&  &  & $\ \searrow ^{S_{E}^{\left( k-2\right) }}$ &  \\ \cline{5-5}
&  &  &  & \multicolumn{1}{|c|}{$\overset{}{\text{Generalized }\mathfrak{bms}%
_{3}\,\,\text{algebra }\,\left(\mathfrak{vir}_{\mathfrak{B}_{k}}\right)}$}
\\ \cline{5-5}
\end{tabular}%
\end{equation*}%
In Sect. 5 we have introduced another family of infinite dimensional
algebras, $\mathfrak{vir}_{\mathfrak{D}_k}$, which can be obtained by
expanding the Virasoro algebra using the semigroup $S_D^{(k-2)}$, and showed
the simplest examples explicitly. These algebras can always be written as
the direct product $\mathfrak{vir}^2\oplus\mathfrak{B}_{k-2}$ after a
suitable change of basis.

In Sect. 6 the Sugawara construction has been applied to expanded Ka\v{c}%
-Moody algebras to obtain the expanded Virasoro algebras and the cases $k=3$
and $k=4$ have been worked out explicitly. This result is remarkable as it
means that these new infinite-dimensional symmetries could be related to some
kind of generalized WZW theories whose current algebras are given by
expanded Ka\v{c}-Moody algebras. In that case the algebras $\mathfrak{vir}_{%
\mathfrak{B}_k}$ or $\mathfrak{vir}_{\mathfrak{C}_k}$ should be recovered as
the Poisson algebras for the stress-energy momentum tensor components in the
very same way as it happens for $\mathfrak{vir}^2$ and $\mathfrak{bms}_3$.

In the context of gravity, upon imposing suitable boundary conditions, the
algebras $\mathfrak{vir}^2$ and $\mathfrak{bms}_3 $ appear as the asymptotic
symmetries of asymptotically $AdS$ and Asymptotically flat three-dimensional
Einstein gravity, respectively. We conjecture that the new infinite
dimensional algebras $\mathfrak{vir}_{\mathfrak{B}_k}$, $\mathfrak{vir}_{%
\mathfrak{C}_k}$ and $\mathfrak{vir}_{\mathfrak{D}_k}$ obtained here
correspond the asymptotic symmetries of 3D gravity theories invariant under
the algebras $\mathfrak{B}_k$, $\mathfrak{C}_k$ or $\mathfrak{D}_k$ when
suitable boundary conditions for the fields content are adopted. These
theories of gravity can be straightforwardly constructed by considering
Chern-Simons actions invariant under these algebras. This will be the
subject of a subsequent article.

On the other hand, it is well-known that the KdV system possesses a Virasoro
symmetry related to the KdV hierarchy \cite{Batlle}. This result can be used
to construct an infinite set of boundary conditions for 3D gravity \cite{PTT}%
. Along this line it would be interesting to evaluate the existence of
integrable systems associated with expanded Virasoro symmetries and they
hierarchies as well as their relations to boundary conditions for gravity
theories invariant under the algebras $\mathfrak{B}_k$ or $\mathfrak{C}_k$.

Another natural generalization of our results is to extend the expansion
method to $\mathcal{N}$-extended supersymmetric extension of asymptotic
symmetries. In particular, it would be interesting to study $S$-expanded
super Virasoro symmetries. However, this would require a more subtle
treatment than the one introduced here. Indeed, one cannot naively consider
the expansion of a super Virasoro structure. The general setup and the
respective supergravity models will be presented in a future paper. As an
ending remark: it would be worth exploring the expansion procedure to higher
spin extension of gravity theories in $2+1$ dimensions.

\section{Acknowledgments}

This work was supported by the Chilean FONDECYT Projects N$^{\circ}$3170437
(P.C.), N$^{\circ}$3170438 (E.R.) and N$^{\circ}$3160581 (P.S-R.). The
authors wish to thank O. Fuentealba, J. Matulich, G. Silva and R. Troncoso
for valuable discussion and comments. R.C. would like to thank to the
Direcci\'on de Investigaci\'on and Vice-rector\'ia de Investigaci\'on of the
Universidad Cat\'olica de la Sant\'isima Concepci\'on, Chile, for their
constant support.

\section*{Appendix}

\appendix

\section{Generalized Ka\v{c}-Moody algebras\label{Appc}}

Let us consider the Ka\v{c}-Moody algebra $\mathfrak{\hat{\mathfrak{g}}}_{k}$%
\thinspace ,
\begin{equation}
\left[ j_{m}^{a},j_{n}^{b}\right] =if_{\quad
c}^{ab}j_{m+n}^{c}+kmg^{ab}\delta _{m,-n}\,,
\end{equation}%
which corresponds to the central extension of the loop algebra of a
semisimple Lie algebra $\mathfrak{g}$. One can show that two families of Ka%
\v{c}-Moody-like algebras can be obtained applying diverse semigroups $S$ to
$\mathfrak{\hat{\mathfrak{g}}}_{k}$. Let $S_{E}^{\left( k-2\right) }=\left\{
\lambda _{0},\lambda _{1},\dots ,\lambda _{k-1}\right\} $ be the finite
abelian semigroup whose elements satisfy (\ref{sek2}) and $\lambda
_{k-1}=0_{s}$ is the zero element of the semigroup. Then, the $S_{E}^{\left(
k-2\right) }$-expanded algebra is given by
\begin{equation}
\left[ j_{\left( m,\alpha \right) }^{a},j_{\left( n,\beta \right) }^{b}%
\right] =\left\{
\begin{array}{lcl}
if_{\quad c}^{ab}j_{\left( m+n,\alpha +\beta \right) }^{c}+k_{\alpha +\beta
+1}mg^{ab}\delta _{m,-n}\,\,\,\, & \mathrm{if}\,\,\,\,\alpha +\beta \leq k-2
&  \\
0\,\,\,\, & \mathrm{if}\,\,\,\,\alpha +\beta >k-2 &
\end{array}%
\right.
\end{equation}%
where we have defined $k_{\alpha +\beta +1}\equiv k_{\alpha \beta
}=kK_{\alpha \beta }^{\gamma }\lambda _{\gamma }$. One can see that the $%
S_{E}^{\left( k-2\right) }$-expanded algebras always contain an abelian
ideal generated by the set
\begin{equation*}
\mathcal{A}=\left\{ j_{\left( m,\tilde{\alpha}\right) }^{a}\right\}
\,\,\,\,,\,\,\,\,\text{$\tilde{\alpha}$}=\left[ \frac{k}{2}\right] ,\ldots
,k-2
\end{equation*}%
and for which
\begin{equation}
\begin{array}{lcl}
\left[ j_{\left( m,\tilde{\alpha}\right) }^{a},j_{\left( m,\tilde{\beta}%
\right) }^{b}\right] & = & 0\,, \\[5pt]
\left[ j_{\left( m,\tilde{\alpha}\right) }^{a},j_{\left( m,\alpha \right)
}^{b}\right] & \in & \mathcal{A}\,\,+\mathrm{central\,\,terms} \\[5pt]
&  &
\end{array}%
\end{equation}%
In particular, the Ka\v{c}-Moody-like structure appears by redefining the
generators in the form
\begin{equation}
\begin{array}{lcl}
j_{m}^{a\left( i\right) } & \equiv & j_{(m,i)}^{a}=\lambda _{i}\ell _{m}\,,
\\[5pt]
p_{m}^{a\left( \bar{\imath}\right) } & \equiv & j_{(m,\bar{\imath}%
)}^{a}=\lambda _{\bar{\imath}}\ell _{m}\,,%
\end{array}
\label{redefkm}
\end{equation}%
where $i$ takes even values and $\bar{\imath}$ takes odd values. This allows
one to write the $S_{E}^{\left( k-2\right) }$-expanded algebras in the form
\begin{equation}
\begin{array}{lcl}
\left[ j_{m}^{a\left( i\right) },j_{n}^{b\left( j\right) }\right] & = &
if_{\quad c}^{ab}j_{m+n}^{c\left( i+j\right) }+k_{i+j+1}mg^{ab}\delta
_{m,-n}\,\,\,\,,\,\,\,\,\mathrm{for}\,\,\,\,i+j\leq k-2\,, \\[5pt]
\left[ j_{m}^{a\left( i\right) },p_{n}^{b\left( \bar{\imath}\right) }\right]
& = & if_{\quad c}^{ab}p_{m+n}^{c\left( i+\bar{\imath}\right) }+k_{i+\bar{%
\imath}+1}mg^{ab}\delta _{m,-n}\,\,\,\,,\,\,\,\,\mathrm{for}\,\,\,\,i+\bar{%
\imath}\leq k-2\,, \\[5pt]
\left[ p_{m}^{a\left( \bar{\imath}\right) },p_{n}^{b\left( \bar{j}\right) }%
\right] & = & if_{\quad c}^{ab}j_{m+n}^{c\left( \bar{\imath}+\bar{j}\right)
}+k_{\bar{\imath}+\bar{j}+1}mg^{ab}\delta _{m,-n}\,\ ,\,\,\ \ \mathrm{for}%
\,\,\,\,\bar{\imath}+\bar{j}\leq k-2\,, \\
\mathrm{others} & = & 0\,.%
\end{array}
\label{kml}
\end{equation}%
Let us note that for $k=3$, the semigroup corresponds to the $S_{E}^{\left(
1\right) }$ whose elements satisfy (\ref{SE1}) and the commutation relations
(\ref{kml}) reduce to the affine current algebra given by\ (\ref{km}). The
case $k=4$ reproduce the $S_{E}^{\left( 2\right) }$-expanded algebra whose
generators satisfy (\ref{defkm}).

An alternative family of generalized Ka\v{c}-Moody algebras can be obtained
applying the $S_{\mathcal{M}}^{\left( k-2\right) }=\left\{ \lambda
_{0},\lambda _{1},\dots ,\lambda _{k-2}\right\} $ semigroup to $\mathfrak{%
\hat{\mathfrak{g}}}_{k}$. Considering the multiplication law of the
semigroup (\ref{smk2}) one can show that the $S_{\mathcal{M}}^{\left(
k-2\right) }$-expanded algebra takes the form
\begin{equation}
\left[ j_{\left( m,\alpha \right) }^{a},j_{\left( n,\beta \right) }^{b}%
\right] =\left\{
\begin{array}{lcl}
if_{\quad c}^{ab}j_{\left( m+n,\alpha +\beta \right) }^{c}+k_{\alpha +\beta
+1}mg^{ab}\delta _{m,-n}\,\,\,\, & \mathrm{if}\,\,\,\,\alpha +\beta \leq k-2
&  \\
if_{\quad c}^{ab}j_{\left( m+n,\alpha +\beta -2\left[ \frac{k-1}{2}\right]
\right) }^{c}+k_{\alpha +\beta -2\left[ \frac{k-1}{2}\right]
+1}mg^{ab}\delta _{m,-n}\,\,\,\, & \mathrm{if}\,\,\,\,\alpha +\beta >k-2 &
\end{array}%
\right.
\end{equation}%
One can redefine the generators in the form (\ref{redefkm}) leading to the
following generalized affine current algebra
\begin{equation}
\begin{array}{lcl}
\left[ j_{m}^{a\left( i\right) },j_{n}^{b\left( j\right) }\right] & = &
if_{\quad c}^{ab}j_{m+n}^{c\left\{ i+j\right\} }+k_{\left\{ i+j\right\}
+1}mg^{ab}\delta _{m,-n}\,, \\[5pt]
\left[ j_{m}^{a\left( i\right) },p_{n}^{b\left( \bar{\imath}\right) }\right]
& = & if_{\quad c}^{ab}p_{m+n}^{c\left\{ i+\bar{\imath}\right\} }+k_{\left\{
i+\bar{\imath}\right\} +1}mg^{ab}\delta _{m,-n}\,, \\[5pt]
\left[ p_{m}^{a\left( \bar{\imath}\right) },p_{n}^{b\left( \bar{j}\right) }%
\right] & = & if_{\quad c}^{ab}j_{m+n}^{c\left\{ \bar{\imath}+\bar{j}%
\right\} }+k_{\left\{ \bar{\imath}+\bar{j}\right\} +1}mg^{ab}\delta
_{m,-n}\,,%
\end{array}
\label{gkm}
\end{equation}%
where $\left\{ \cdots \right\} $ means the following%
\begin{equation}
\left\{ i+j\right\} =\left\{
\begin{array}{lcl}
i+j & \mathrm{if}\,\,\,\,i+j\leq k-2 &  \\
i+j-2\left[ \frac{k-1}{2}\right] \, & \mathrm{if}\,\,\,\,i+j>k-2 &
\end{array}%
\right.
\end{equation}%
Interestingly the $k=3$ and $k=4$ cases reproduce two and three copies of Ka%
\v{c}-Moody algebras, respectively. However, for $k\geq 5$ the commutation
relations of the generalized Ka\v{c}-Moody algebra obtained here become
non-trivial and are given by (\ref{gkm}).

It is important to mention that the two families of generalized Ka\v{c}%
-Moody algebras presented here are related through the IW contraction.
Indeed, considering the rescaling of the generators satisfyng a $S_{\mathcal{%
M}}^{\left( k-2\right) }$-expanded algebra (\ref{gkm})%
\begin{eqnarray*}
j_{m}^{0} &\rightarrow &j_{m}^{0}\,,\text{ }j_{m}^{i}\rightarrow \sigma
^{i}j_{m}^{i}\text{ },p_{m}^{\bar{\imath}}\rightarrow \sigma ^{\bar{\imath}%
}p_{m}^{\bar{\imath}}\,,\text{ } \\
\text{ }k_{1} &\rightarrow &k_{1}\,,\text{ }k_{i+1}\rightarrow \sigma
^{i}k_{i+1}\,,\text{ }k_{\bar{\imath}+1}\rightarrow \sigma ^{\bar{\imath}}k_{%
\bar{\imath}+1}
\end{eqnarray*}%
the limit $\sigma \rightarrow \infty $ leads to the $S_{E}^{\left(
k-2\right) }$-expanded algebra (\ref{kml}).

{\small {}{}\bigskip{} }

{\small {}{}}

\end{document}